\documentclass[%
 reprint,
superscriptaddress,
 amsmath,amssymb,
 aps,
prx,
longbibliography,
]{revtex4-2}

\usepackage{graphicx}
\usepackage{dcolumn}
\usepackage{bm}

\usepackage{color}

\usepackage{amsmath}
\usepackage{bm}
\usepackage{lmodern}
\usepackage{siunitx}
\usepackage{booktabs}
\usepackage{etoolbox}
\usepackage{tabularx}
\usepackage{booktabs}
\usepackage{multirow}
\usepackage{threeparttable}
\usepackage{physics}
\usepackage{url}
\usepackage{enumerate}
\usepackage{autobreak}
\usepackage{lipsum}

\usepackage[
    hidelinks
]{hyperref}

\newcommand{\RomanNumeralCaps}[1]
    {\MakeUppercase{\romannumeral #1}}



\newcommand{\mycomment}[1]{}

\begin{document}

\preprint{APS/123-QED}

\title{Generalizing Abell--Tersoff bond-order potential with explicit high-order many-body correlations for robust extrapolation of potential energy surfaces}

\author{Ikuma Kohata}
\affiliation{Department of Materials Engineering, The University of Tokyo, 7-3-1 Hongo, Bunkyo-ku, Tokyo 113-8656, Japan}

\date{\today}

\begin{abstract}
Machine-learning interatomic potentials enable accurate and efficient atomistic simulations, yet their reliability for out-of-distribution configurations far beyond the training domain remains a significant challenge. Here, we introduce a semiparametric interatomic potential based on a generalization of the Abell--Tersoff bond-order potential, incorporating a chemically informed functional form and explicit high-order many-body correlations to address this challenge. The model is trained and evaluated on various datasets of silicon, carbon, water, and small molecules, achieving interpolation accuracy comparable to existing MLIP models while exhibiting improved extrapolation to unseen configurations, including those at high pressures and temperatures. These results provide insights into the design of specific inductive biases for reliable extrapolation in interatomic potentials.
\end{abstract}

\maketitle

\section{Introduction}
Under the Born–Oppenheimer approximation, the potential energy of a system containing interacting atoms can be represented as a function of their nuclear positions and chemical species. This relationship, referred to as the Born–Oppenheimer potential energy surface, plays an important role in a wide range of problems in physics, chemistry, and materials science. A wide variety of methods have been proposed to describe potential energy surfaces, each having different levels of accuracy and computational cost. Although first-principles calculations based on density functional theory (DFT) provide highly accurate potential energy surfaces and have been widely used, their high computational cost arising from the explicit treatment of electrons limits the size and timescale of atomistic simulations.

To overcome this limitation, a variety of approximations to potential energy surfaces have been developed. In particular, interatomic potentials have been widely used for efficient atomistic simulations. By bypassing explicit electronic-structure calculations and exploiting the locality of interatomic interactions, they enable fast evaluation of potential energies and forces, thereby allowing large-scale atomistic simulations at a fraction of the computational cost of first-principles methods. Over the past several decades, many forms of interatomic potentials have been proposed, which can broadly be classified into two categories: empirical interatomic potentials and machine-learning interatomic potentials (MLIPs).

Empirical interatomic potentials are parametric approximations of potential energy surfaces, constructed using predefined functional forms motivated by chemical intuition. Widely known examples are the embedded atom method (EAM) \cite{PhysRevB.29.6443}, the potential proposed by Stillinger and Weber \cite{PhysRevB.31.5262}, and the bond-order potential formulated by Abell \cite{PhysRevB.31.6184} and Tersoff \cite{PhysRevLett.56.632,PhysRevB.38.9902}. Their chemically motivated functional forms can impose physically reasonable behavior beyond the training domain. However, their limited flexibility often prevents accurate predictions across diverse atomic configurations, thereby constraining their applicability.

In contrast, MLIPs are nonparametric approximations of potential energy surfaces constructed using machine-learning models with a large number of trainable parameters. Since the early studies introducing Behler–Parrinello-type neural network potential \cite{PhysRevLett.98.146401} and Gaussian Approximation Potential (GAP) \cite{PhysRevLett.104.136403}, a wide range of MLIP architectures and applications have been explored. Their flexibility enables accurate representation of complex potential energy surfaces beyond the accuracy achievable with conventional empirical interatomic potentials, allowing high-fidelity simulations across diverse configurations when sufficient training data are available. However, such flexibility comes at the cost of reduced robustness; MLIPs generally exhibit poor extrapolation beyond the training domain \cite{Benoit_2021,Vita_2023}, which often results in unphysical behavior during molecular dynamics (MD) simulations \cite{fu2023forces,10.1063/5.0218705,Cui2025}. This trade-off between accuracy and robustness has been a challenge in developing reliable and data-efficient MLIPs.

To improve the accuracy and transferability of MLIPs, the development of local-environment descriptors has played a central role. In particular, systematically improvable body-ordered representations of local atomic environments, such as the atomic cluster expansion (ACE) \cite{PhysRevB.99.014104} and moment tensor potential (MTP) \cite{doi:10.1137/15M1054183}, have improved both accuracy and transferability by reducing degeneracies in local-environment representations through the inclusion of explicit higher-order correlations \cite{PhysRevLett.125.166001}. However, although these frameworks provide complete basis expansions with well-defined body-ordered structures, the individual basis functions possess limited physical interpretability beyond their body order, and robust prediction for atomic configurations far away from the training domain is still a difficult task.

Recent studies on foundation MLIPs \cite{Chen2022,Takamoto2022,D2DD00096B,Merchant2023,Deng2023,D2DD00096B,yang2024mattersimdeeplearningatomistic,neumann2024orbfastscalableneural,doi:10.1021/acs.jctc.4c00190,barrosoluque2026openmaterials2024omat24,10.1063/5.0297006,Lysogorskiy2026}, which are large MLIP architectures trained on massive datasets, have demonstrated that they can achieve high accuracy across diverse configurational and chemical spaces. Nevertheless, they are also subject to the same limitation: reduced accuracy for unusual configurations that are rarely present in or entirely absent from the training data, such as highly compressed structures. As demonstrated by the homonuclear diatomic energy curves reported in the MLIP Arena benchmark \cite{chiang2026mlip}, they often exhibit unphysical artifacts on potential energy surfaces, including spurious local energy minima, false energy barriers, and oscillatory behavior. Thus, beyond achieving low regression errors, ensuring physically reasonable behavior throughout configuration space remains a challenge in high-quality modeling of potential energy surfaces.

For a more reliable description of potential energy surfaces, several studies have developed interatomic potential models that combine the parametric forms of empirical interatomic potentials with generic regression methods, such as neural networks and symbolic regression \cite{Pun2019,73gp-bm46,Hernandez2019,PhysRevMaterials.7.053804,Varughese2026,PhysRevB.102.144107}. Among these approaches, bond-order-based MLIPs \cite{Pun2019,73gp-bm46}, which combine the chemical intuition of bond-order potentials with the flexibility of neural-network regression, have demonstrated improved robustness in extrapolative regions of the potential energy surface compared with purely nonparametric MLIPs. However, their neural-network-based formulations compromise physical interpretability, and explicit functional forms required for reliable extrapolation are not fully understood. Furthermore, their applicability to multi-element systems has not been demonstrated, and their extrapolation capability has been less systematically evaluated on publicly available datasets.

In this study, motivated by these considerations, we introduce a semiparametric interatomic potential based on a body-ordered extension of the Abell--Tersoff bond-order potential, analogous to how ACE and MTP provide body-ordered extensions of cluster potentials such as the Stillinger--Weber potential \cite{PhysRevB.31.5262} and the Biswas--Hamann potential \cite{PhysRevLett.55.2001,PhysRevB.36.6434}. We refer to this potential as the body-ordered bond-order potential (BOBOP). BOBOP retains the analytical functional form and chemical interpretability of Abell--Tersoff bond-order potentials while explicitly incorporating high-order many-body correlations, and is readily applicable to multi-element systems. BOBOP is trained and evaluated on a range of publicly available datasets spanning silicon, carbon, water, and small molecules, demonstrating robust extrapolation to out-of-distribution configurations compared with existing local MLIP models.

The remainder of this paper is organized as follows. Section~\ref{sec:background} reviews the conventional Abell--Tersoff bond-order potential and its limitations. Section~\ref{sec:method} introduces BOBOP as a body-ordered extension of the Abell--Tersoff bond-order potential. Section~\ref{sec:asymptotic} describes the asymptotic behavior of BOBOP and its bond order in extrapolation domains. Section~\ref{sec:highorder} investigates the effect of explicit high-order correlations in BOBOP on accuracy and data efficiency.
Section~\ref{sec:training} presents the application of BOBOP to various training datasets and evaluates its interpolation accuracy and extrapolation robustness in comparison with existing MLIP models. Section~\ref{sec:discussion} discusses the key components of BOBOP contributing to the accuracy and transferability. Finally, Section~\ref{sec:conclusion} summarizes this study.

\section{Background}
\label{sec:background}

The original concept of the bond-order potential was proposed by Abell \cite{PhysRevB.31.6184}, who expressed the local attractive interaction contributing to the binding energy as
\begin{equation}
E_{i}^\mathrm{el} = -\sum_{j \neq i} b_{ij}V_{A}(r_{ij}),
\end{equation}
where $i$ indexes atoms in the system, $j$ indexes neighboring atoms around atom $i$, $r_{ij}$ is the distance between atom $i$ and atom $j$, $V_{A}$ is the pairwise attractive term, and $b_{ij}$ is the bond-order function describing many-body effects on bond $i$-$j$. Based on a Bethe lattice analysis, Abell suggested that the bond order is primarily determined by the coordination number $Z$, and approximated the bond order as 
\begin{equation}
\label{eq:abell}
b_{ij} = Z^{-\delta},
\end{equation}
where $\delta$ is an adjustable parameter typically restricted to be positive. This relationship describes the behavior of bond saturation: as the coordination number increases, a limited number of valence electrons are distributed over a larger number of bonds, leading to a reduction in the bond strength of each bond.

The practical formulation of the bond-order potential was later developed by Tersoff \cite{PhysRevLett.56.632,PhysRevB.38.9902}.
In the Abell-Tersoff bond-order potential, the cohesive energy of a system $E_\mathrm{coh}$ is given by
\begin{align}
E_\mathrm{coh} &= \sum_{i}E_{i}, \\
E_{i} &= \sum_{j \neq i}V_{ij}, \\
V_{ij} &= f(r_{ij})\bigl[V_{R}(r_{ij})-b_{ij}V_{A}(r_{ij})\bigr],
\end{align}
where $V_{R}$ is the pairwise repulsive term, and $f_{c}$ is the cutoff function that smoothly truncates the interaction at the cutoff distance.
The bond order $b_{ij}$ is given by
\begin{align}
b_{ij} &= \Phi(\zeta_{ij}), \\
\zeta_{ij} &= \sum_{k \neq i,j} f(r_{ik})\exp\bigl[\lambda(r_{ij}-r_{ik})^{m}\bigr] g(\theta_{ijk}),
\label{eq:zeta_bo}
\end{align}
where the index $k$ runs over the neighboring atoms of atom $i$, $\theta_{ijk}$ is the bond angle between bonds $i$-$j$ and $i$-$k$, $g(\theta_{ijk})$ is a function describing the angular dependence, and $\lambda$ is an adjustable parameter. The distance-dependent factor $\exp\bigl[\lambda(r_{ij}-r_{ik})^{m}\bigr]$ describes the competition between bonds $i$-$j$ and $i$-$k$ \cite{PhysRevB.37.6991}; when bond $i$-$j$ is much shorter than bond $i$-$k$, the presence of bond $i$-$k$ has little effect on bond $i$-$j$, whereas bond $i$-$k$ is more strongly weakened.
The parameter $m$ controls the strength of this distance-dependent effect and is typically fixed to $m=3$ \cite{PhysRevLett.56.632,PhysRevB.38.9902} or $m=1$ \cite{PhysRevB.42.9458,PhysRevB.66.035205}. The function $\Phi$ represents the nonlinear relationship between $\zeta_{ij}$ and $b_{ij}$. A general form of $\Phi$ \cite{KUMAGAI2007457} is given by
\begin{align}
    \hat{\zeta}_{ij} &= \zeta_{ij}^{\eta}, \label{eq:bohatzeta} \\
    b_{ij} &= (1+\hat{\zeta}_{ij})^{-\delta}, \label{eq:bobij}
\end{align}
where $\delta$ and $\eta$ are positive adjustable parameters. The quantity $1+\hat{\zeta}_{ij}$ is analogous to the coordination number $Z$ in Eq.~\eqref{eq:abell}, since both increase as the number of neighboring atoms increases.

The bond-order potential was fitted to the physical properties of group-\RomanNumeralCaps{4} elements, such as silicon \cite{PhysRevLett.56.632,PhysRevB.38.9902,PhysRevB.37.6991}, carbon \cite{PhysRevLett.61.2879}, and their multi-element systems \cite{PhysRevB.39.5566,PhysRevB.41.3248.2,PhysRevLett.64.1757}, demonstrating its capability to describe covalent bonding materials. However, it has limited flexibility in representing potential energy surfaces over a wide range of configurations due to several reasons.

One reason is that the local environment is represented by a single scalar quantity $\zeta_{ij}$, whereas modern descriptors of local environments are typically represented as high-dimensional vectors. This scalar representation cannot fully distinguish diverse bonding environments, which limits the attainable accuracy. 

\begin{figure}[tb]
\raggedleft
\includegraphics[clip,scale=0.40]{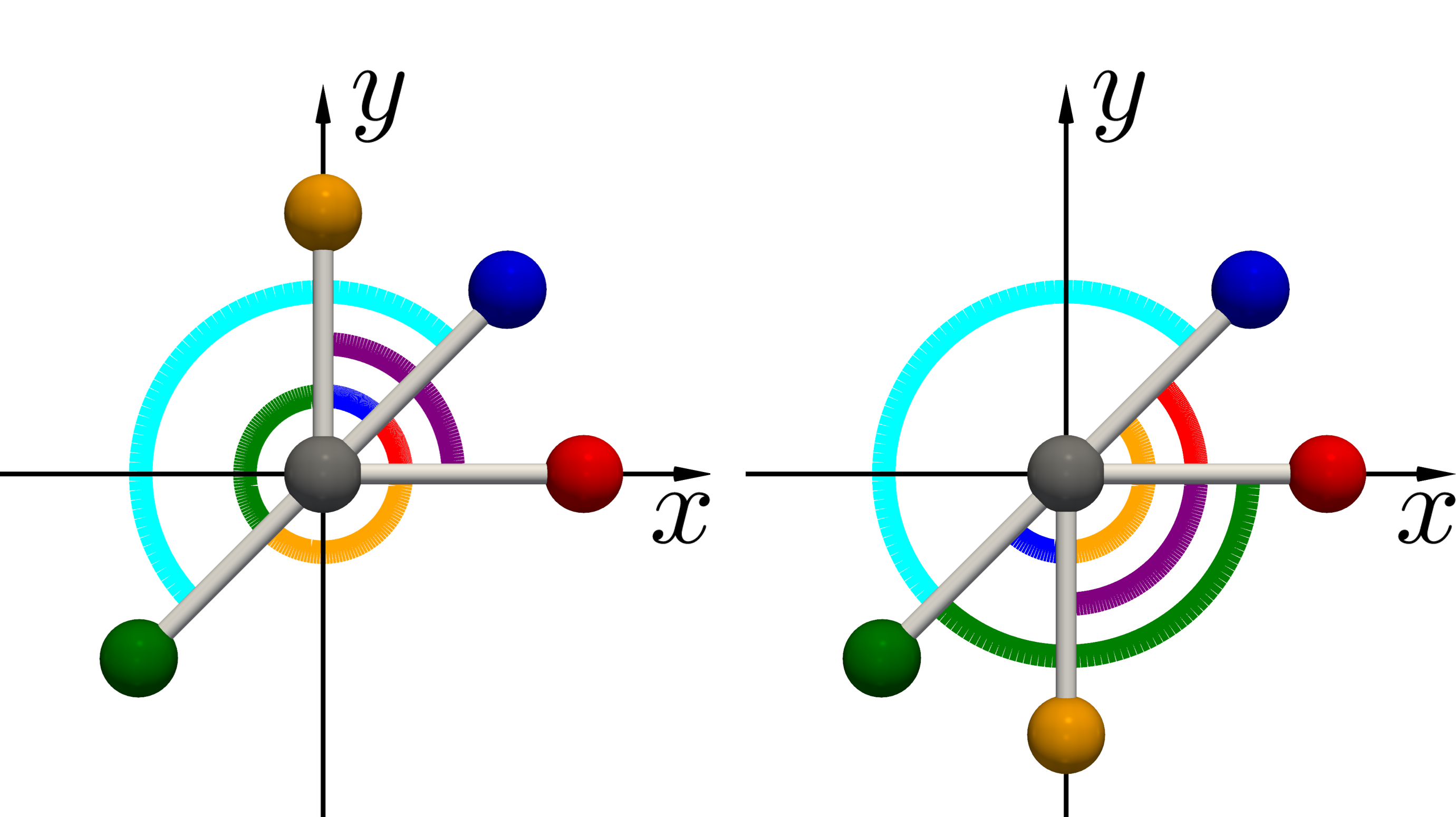}
\caption{Two distinct bonding environments in planar CH$_{4}$ that yield the same value of bond order $b_{ij}$ in the bond-order potential. The gray, red, blue, green, and yellow spheres represent atoms $i$, $j$, $k$, $h$, and $u$. Atom $i$ is a carbon atom. Atoms $j$, $k$, $h$, and $u$ are hydrogen atoms, which are indistinguishable under permutation. All bonds connected to atom $i$ have the same length. The blue, red, purple, yellow, green, and cyan arcs indicate bond angles of 45$^\circ$, 45$^\circ$, 90$^\circ$, 135$^\circ$, 135$^\circ$, and 180$^\circ$, respectively.}
\label{fig:molecule}
\end{figure}

Another reason is the presence of bond-energy degeneracy, which is analogous to the atomic-energy degeneracy \cite{PhysRevLett.125.166001}. In bond-order potentials, the bond order is determined from a nonlinear transformation of three-body terms that depend on bond angles with respect to the bond $i$--$j$, which implicitly introduces higher-order correlations. For example, when $\eta=2$, Eq.~\eqref{eq:bohatzeta} can be expanded as
\begin{align}
\label{eq:zetaexpand}
\hat{\zeta}_{ij} =& \left\{\sum_{k \neq i,j} f(r_{ik})\exp\bigl[\lambda(r_{ij}-r_{ik})^{m}\bigr]g(\theta_{ijk})\right\}^{2} \notag \\
=& \sum_{k \neq i,j}\sum_{h \neq i,j}\left\{f(r_{ik})f(r_{ih})\exp\bigl[\lambda(r_{ij}-r_{ik})^{m}\bigr] \right. \notag \\ & \qquad \qquad \times  \left. \exp\bigl[\lambda(r_{ij}-r_{ih})^{m}\bigr]g(\theta_{ijk})g(\theta_{ijh})\right\}.
\end{align}
This expansion describes many-body correlations between multiple neighboring atoms around the bond $i$--$j$. However, because they are constructed only from three-body angular information relative to the bond $i$--$j$, distinct bonding environments with identical bond-angle histograms can become indistinguishable. Figure~\ref{fig:molecule} illustrates two distinct bonding environments in CH$_{4}$ that yield the same bond order $b_{ij}$. Although the two configurations differ geometrically, they have identical bond-angle histograms relative to the bond $i$--$j$, each containing angles of 45$^\circ$, 90$^\circ$, and 135$^\circ$. Consequently, the two bonding environments are indistinguishable by the bond order $b_{ij}$, resulting in the same bond energy $V_{ij}$. However, if the bond order is intended to provide a chemically meaningful representation of the bonding environment, it should distinguish between these two environments. The bond-order-based MLIPs \cite{Pun2019,73gp-bm46} exhibit the same bond-energy degeneracy because they also rely on the angle histogram relative to the bond $i$--$j$ and atom-centered three-body descriptors, both of which are identical for these two geometries. This degeneracy introduces an ambiguity in the representation of the bonding environment, limiting the physically reasonable description of the bond order.

\mycomment{

It should be noted that this bond-energy degeneracy is analogous to, but distinct from, the atomic-energy degeneracy demonstrated in Ref. \cite{PhysRevLett.125.166001}. In the Abell--Tersoff bond-order potential, the bond energies are degenerate for the two distinct CH$_4$ configurations. However, the atomic energy $E_{i}$, obtained by summing the bond energies $V_{ij}$ over neighboring atoms, is not degenerate, in contrast to atom-centered MLIPs based on three-body descriptors, which exhibit atomic-energy degeneracy for the same configurations \cite{PhysRevLett.125.166001}. In fact, in Fig. \ref{fig:molecule}, the histograms of bond angles relative to the bond $i$--$j$ are identical, containing angles of (45$^\circ$, 90$^\circ$, 135$^\circ$), resulting in the same bond order $b_{ij}$. By contrast, the angle histogram relative to the bond $i$--$k$ in the left configuration consists of (45$^\circ$, 45$^\circ$, 180$^\circ$), whereas no bond in the right configuration has the same angle histogram. As a result, although the Abell--Tersoff potential exhibits bond-energy degeneracy, it distinguishes the atom-centered local environments and therefore does not exhibit atomic-energy degeneracy. This reflects the fact that, despite employing only three-body angular information for each bond, the atomic energy obtained by summing bond energies contains effective four-body correlations. The effective body order of bond-order potentials and its relation to ACE are discussed in Appendix~\ref{sec:relationship}.
}

\section{Body-ordered extension of bond-order potential}
\label{sec:method}
To overcome the limitations discussed in Sec.~\ref{sec:background}, we introduce BOBOP as a body-ordered extension of the bond-order potential.
\subsection{Formulation}

In BOBOP, the total energy is defined using a combination of the repulsion term $V_{R}$, the attraction term $V_{A}$, and the bond-order $b_{ij}$: 
\begin{align}
E_\mathrm{tot} &= \sum_{i}E_{i}, \\
E_{i} &= \sum_{j \neq i}V_{ij} + \varepsilon(\mu_{i})\\
V_{ij} &= f(r_{ij})[V_{R}(r_{ij}) - b_{ij}V_{A}(r_{ij})],
\end{align}
where $E_{i}$ is the site energy of atom $i$, $\mu_{i}$ is the chemical species of atom $i$, and $\varepsilon(\mu_{i})$ is the isolated atom energy specific to $\mu_{i}$. The repulsive term $V_{R}$ and attractive term $V_{A}$ are represented by Morse-type functional forms
\begin{align}
V_{R}(r_{ij}) &= \left(1+\frac{Q_{\mu_{i}\mu_{j}}}{r_{ij}}\right)D_{\mu_{i}\mu_{j}}^{(R)}\exp(-\alpha_{\mu_{i}\mu_{j}}r_{ij}), \\
V_{A}(r_{ij}) &= D_{\mu_{i}\mu_{j}}^{(A)}\exp(-\beta_{\mu_{i}\mu_{j}}r_{ij}), 
\end{align}
where $Q_{\mu_{i}\mu_{j}}$, $D_{\mu_{i}\mu_{j}}^{(R)}$, $D_{\mu_{i}\mu_{j}}^{(A)}$, $\alpha_{\mu_{i}\mu_{j}}$, and $\beta_{\mu_{i}\mu_{j}}$ are trainable parameters depending on the chemical species $\mu_{i}$ and $\mu_{j}$. The cutoff function $f(r)$ is defined as 
\begin{equation} 
f(r) = \left [  \dfrac{\tanh \{ B_\mathrm{cut}(1-r/R_\mathrm{cut}) \} }{\tanh(B_\mathrm{cut})} \right ]^{3} \; (r<R_\mathrm{cut}),
\label{eq:cutoff}
\end{equation}
where $R_\mathrm{cut}$ is the cutoff distance and $B_\mathrm{cut}$ is a trainable parameter. The cutoff function and their first and second derivative vanish at the cutoff distance $R_{c}$. 
The bond order $b_{ij}$ is constructed from the local environment through several steps, described in the following.
\subsubsection{Construction of atom basis}
The edge basis $\phi_{ijn\bm{l}}$, used to describe the spatial arrangement of edge $i$-$j$, is constructed as a product of the cutoff function $f(r_{ij})$, the element-dependent parameter $\chi_{n\mu_{i}\mu_{j}}$, the radial basis $R_{n}(r_{ij})$, and the angular basis $L_{\bm{l}}(\hat{\bm{r}}_{ij})$:
\begin{align} 
\label{eq:edgebasis}
\phi_{ijn\bm{l}} &=  f(r_{ij})\chi_{n\mu_{i}\mu_{j}}R_{n}(r_{ij})L_{\bm{l}}(\hat{\bm{r}}_{ij}),
\end{align}
where $n$ indexes the channel of radial basis, $\hat{\bm{r}}_{ij}$ is the unit vector corresponding to edge $i$-$j$, defined as $\hat{\bm{r}}_{ij} \equiv (\hat{x}_{ij},\hat{y}_{ij},\hat{z}_{ij})$. The vector $\bm{l} \equiv (l_{x}, l_{y}, l_{z})$ is a set of angular momentum numbers satisfying $l_{x}+l_{y}+l_{z}=l$, where $l$ denotes the total angular momentum. The angular basis $L_{\bm{l}}(\hat{\bm{r}}_{ij})$ is represented by
\begin{equation} 
L_{\bm{l}}(\hat{\bm{r}}_{ij}) = \hat{x}_{ij}^{l_{x}}\hat{y}_{ij}^{l_{y}}\hat{z}_{ij}^{l_{z}}.
\end{equation}
This angular basis is related to the bond angle through the multinomial theorem:
\begin{equation}
\label{eq:addition}
\cos^{l}\theta_{ijk} = \sum_{\bm{l}}C(\bm{l})L_{\bm{l}}(\hat{\bm{r}}_{ij})L_{\bm{l}}(\hat{\bm{r}}_{ik}),
\end{equation}
where $C(\bm{l})$ is the multinomial coefficient, defined as $C(\bm{l})=l!/(l_{x}!l_{y}!l_{z}!)$. 
The radial basis $R_{n}(r_{ij})$ is given by
\begin{equation}
R_{n}(r_{ij}) = \exp(-\lambda_{n}r_{ij}),
\end{equation}
where $\lambda_{n}$ is a positive trainable parameter corresponding to channel $n$.
The parameters $\chi_{n\mu_{i}\mu_{j}}$ differentiate the edge chemical types and are specific to chemical species $\mu_{i},\mu_{j}$ and channel $n$. The index $n$ takes values of $1,\ldots,n_\mathrm{max}$ and the total angular momentum $l$ takes $0,\ldots,l_\mathrm{max}$.
The atom basis $A_{in\bm{l}}$ is obtained by summing $\phi_{ijn\bm{l}}(\bm{r}_{ij})$ over all edges connecting to atom $i$:
\begin{equation} 
\label{eq:aggregation}
A_{in\bm{l}} = \sum_{j \neq i} \phi_{ijn\bm{l}}.
\end{equation}
\subsubsection{Three-body feature}
Rotationally invariant edge features are constructed as linear combinations of products of $A_{in\bm{l}}$ and $L_{\bm{l}}(\hat{\bm{r}}_{ij})$, with coefficients required for symmetrization. The three-body features $B^{(3)}$ are obtained as 
\begin{align} 
\label{eq:threebody}
B_{ijnl}^{(3)} &= \frac{\sum_{\bm{l}}C(\bm{l})L_{\bm{l}}(\hat{\bm{r}}_{ij})A_{in\bm{l}}}{R_{n}(r_{ij})} \notag \\
&= \sum_{k \neq i}\chi_{n\mu_{i}\mu_{k}}f(r_{ik})\exp\left[\lambda_{n}(r_{ij}-r_{ik})\right]\cos^{l}\theta_{ijk}.
\end{align}
These features contain the self-interaction term corresponding to $k=j$ and therefore remain nonzero even when bond $i$-$j$ has no neighboring atoms.
To remove this self-interaction, the $k=j$ contribution is subtracted from $B^{(3)}$ as 
\begin{align}
&\Tilde{B}_{ijnl}^{(3)} \notag \\ 
&= B_{ijnl}^{(3)} - f(r_{ij})\chi_{n\mu_{i}\mu_{j}} \notag \\
&= \sum_{k \neq i,j}\chi_{n\mu_{i}\mu_{k}}f(r_{ik})\exp\left[\lambda_{n}(r_{ij}-r_{ik})\right]\cos^{l}\theta_{ijk},
\end{align}
where $\Tilde{B}^{(3)}$ denotes the non-self-interacting three-body features, which become zero when bond $i$-$j$ has no neighboring atoms.
Taking a linear combination of $\Tilde{B}^{(3)}$ over $l$ recovers Eq.~\eqref{eq:zeta_bo} when $m=1$, $\chi_{n\mu_{i}\mu_{k}}=1$, $\lambda_{n}=\lambda$, and $g(\theta_{ijk})$ is expanded as a polynomial in  $\cos\theta_{ijk}$. 
\subsubsection{Four-body feature}
The four-body features $B^{(4)}$ are constructed as 
\begin{widetext}
\begin{align} 
B_{\substack{ijn_{1}n_{2}\\l_{1}l_{2}l_{3}}}^{(4)} &= \frac{\sum_{\bm{l}_{1}\bm{l}_{2}\bm{l}_{3}}C(\bm{l}_{1})C(\bm{l}_{2})C(\bm{l}_{3})L_{(\bm{l}_{1}+\bm{l}_{2})}(\hat{\bm{r}}_{ij}) A_{in_{1}(\bm{l}_{1}+\bm{l}_{3})}A_{in_{2}(\bm{l}_{2}+\bm{l}_{3})}}{R_{n_{1}}(r_{ij})R_{n_{2}}(r_{ij})} \notag \\
&= \sum_{k \neq i}\sum_{h \neq i}\chi_{n_{1}\mu_{i}\mu_{k}}\chi_{n_{2}\mu_{i}\mu_{h}}
f(r_{ik})f(r_{ih})\exp\left[\lambda_{n_{1}}(r_{ij}-r_{ik})\right]\exp\left[\lambda_{n_{2}}(r_{ij}-r_{ih})\right]\cos^{l_{1}}\theta_{ijk}\cos^{l_{2}}\theta_{ijh}\cos^{l_{3}}\theta_{ikh}.
\end{align}
As in the three-body case, the self-interaction contributions corresponding to $k=j$ and $h=j$ are removed as
\begin{align}
\label{eq:B4}
\Tilde{B}_{\substack{ijn_{1}n_{2}\\l_{1}l_{2}l_{3}}}^{(4)}
&= B_{\substack{ijn_{1}n_{2}\\l_{1}l_{2}l_{3}}}^{(4)} - f(r_{ij})\left[\Tilde{B}_{ijn_{1}(l_{1}+l_{3})}^{(3)}\chi_{n_{2}\mu_{i}\mu_{j}}+\Tilde{B}_{ijn_{2}(l_{2}+l_{3})}^{(3)}\chi_{n_{1}\mu_{i}\mu_{j}}\right] -  f(r_{ij})^2\chi_{n_{1}\mu_{i}\mu_{j}}\chi_{n_{2}\mu_{i}\mu_{j}} \notag \\
&= \sum_{k \neq i,j}\sum_{h \neq i,j}\chi_{n_{1}\mu_{i}\mu_{k}}\chi_{n_{2}\mu_{i}\mu_{h}}
f(r_{ik})f(r_{ih})\exp\left[\lambda_{n_{1}}(r_{ij}-r_{ik})\right]\exp\left[\lambda_{n_{2}}(r_{ij}-r_{ih})\right]\cos^{l_{1}}\theta_{ijk}\cos^{l_{2}}\theta_{ijh}\cos^{l_{3}}\theta_{ikh},
\end{align}
where $\Tilde{B}^{(4)}$ denotes the corrected four-body features, which become zero when bond $i$-$j$ has no neighboring atoms. These four-body features enable the explicit incorporation of correlations between neighboring bonds $i$-$k$ and $i$-$h$, including the bond angle $\theta_{ikh}$, which cannot be captured by three-body features alone. It should be noted that the present self-interaction correction removes only the contributions of $k=j$ and $h=j$, while the contributions with $k=h$ are retained. Thus, $\Tilde{B}^{(4)}$ are not purely four-body features and still contain three-body contributions, which implies that they remain nonzero even when bond $i$--$j$ has only one neighboring atom. Removing the $k=h$ contributions would require additional computational cost and is therefore not pursued in the present work. 
\subsubsection{Five-body feature}
The five-body features $B^{(5)}$ are obtained as
\begin{align} 
B_{\substack{ijn_{1}n_{2}n_{3}\\l_{1}l_{2}l_{3}l_{4}l_{5}}}^{(5)} &= \frac{\sum_{\bm{l}_{1}\bm{l}_{2}\bm{l}_{3}\bm{l}_{4}\bm{l}_{5}}C(\bm{l}_{1})C(\bm{l}_{2})C(\bm{l}_{3})C(\bm{l}_{4})C(\bm{l}_{5}) L_{(\bm{l}_{1}+\bm{l}_{2}+\bm{l}_{3})}(\hat{\bm{r}}_{ij}) A_{in_{1}(\bm{l}_{1}+\bm{l}_{4})} A_{in_{2}(\bm{l}_{2}+\bm{l}_{4}+\bm{l}_{5})}A_{in_{3}(\bm{l}_{3}+\bm{l}_{5})}}{R_{n_{1}}(r_{ij})R_{n_{2}}(r_{ij})R_{n_{3}}(r_{ij})} \notag \\
&= \sum_{k \neq i,j}\sum_{h \neq i,j}\sum_{u \neq i,j}\chi_{n_{1}\mu_{i}\mu_{k}}\chi_{n_{2}\mu_{i}\mu_{h}}\chi_{n_{3}\mu_{i}\mu_{u}} f(r_{ik})f(r_{ih})f(r_{iu})\exp\left[\lambda_{n_{1}}(r_{ij}-r_{ik})\right]\notag \\ & \quad \times \exp\left[\lambda_{n_{2}}(r_{ij}-r_{ih})\right]\exp\left[\lambda_{n_{3}}(r_{ij}-r_{iu})\right] \cos^{l_{1}}\theta_{ijk}\cos^{l_{2}}\theta_{ijh}\cos^{l_{3}}\theta_{iju}\cos^{l_{4}}\theta_{ikh}\cos^{l_{5}}\theta_{ihu}.
\end{align}
The self-interaction correction is applied as 
\begin{align}
\Tilde{B}_{\substack{ijn_{1}n_{2}n_{3}\\l_{1}l_{2}l_{3}l_{4}l_{5}}}^{(5)} =& B_{\substack{ijn_{1}n_{2}n_{3}\\l_{1}l_{2}l_{3}l_{4}l_{5}}}^{(5)}  \notag \\
&- f(r_{ij})\left[\chi_{n_{3}\mu_{i}\mu_{j}}\Tilde{B}_{ijn_{1}n_{2}l_{1}(l_{2}+l_{5})l_{4}}^{(4)} + \chi_{n_{2}\mu_{i},\mu_{j}}\Tilde{B}_{ijn_{1}(l_{1}+l_{4})}^{(3)}\Tilde{B}_{ijn_{3}(l_{3}+l_{5})}^{(3)}+\chi_{n_{1}\mu_{i}\mu_{j}}\Tilde{B}_{ijn_{2}n_{3}(l_{2}+l_{4})l_{3}l_{5}}^{(4)}\right] \notag \\ 
&- f(r_{ij})^{2}\left[\chi_{n_{2}\mu_{i}\mu_{j}}\chi_{n_{3}\mu_{i}\mu_{j}}\Tilde{B}_{ijn_{1}(l_{1}+l_{4})}^{(3)}+\chi_{n_{1}\mu_{i}\mu_{j}}\chi_{n_{3}\mu_{i}\mu_{j}}\Tilde{B}_{ijn_{2}(l_{2}+l_{4}+l_{5})}^{(3)} +\chi_{n_{1}\mu_{i}\mu_{j}}\chi_{n_{2}\mu_{i}\mu_{j}}\Tilde{B}_{ijn_{3}(l_{3}+l_{5})}^{(3)}\right] \notag \\
&- f(r_{ij})^{3}\chi_{n_{1}\mu_{i}\mu_{j}}\chi_{n_{2}\mu_{i}\mu_{j}}\chi_{n_{3}\mu_{i}\mu_{j}} \notag \\
=& \sum_{k \neq i,j}\sum_{h \neq i,j}\sum_{u \neq i,j}\chi_{n_{1}\mu_{i}\mu_{k}}\chi_{n_{2}\mu_{i}\mu_{h}}\chi_{n_{3}\mu_{i}\mu_{u}} f(r_{ik})f(r_{ih})f(r_{iu})\exp\left[\lambda_{n_{1}}(r_{ij}-r_{ik})\right]\notag \\ & \quad \times \exp\left[\lambda_{n_{2}}(r_{ij}-r_{ih})\right]\exp\left[\lambda_{n_{3}}(r_{ij}-r_{iu})\right] \cos^{l_{1}}\theta_{ijk}\cos^{l_{2}}\theta_{ijh}\cos^{l_{3}}\theta_{iju}\cos^{l_{4}}\theta_{ikh}\cos^{l_{5}}\theta_{ihu},
\label{eq:lincomb}
\end{align}
where $\Tilde{B}^{(5)}$ denotes the corrected five-body features, which become zero when the bond $i$--$j$ has no neighboring atoms. As in the four-body case, the present self-interaction correction removes only the self-interaction terms corresponding to $k=j$, $h=j$, and $u=j$, while the contributions arising from $k=h$, $h=u$, and $u=k$ are retained, which means that $\Tilde{B}^{(5)}$ are also not purely five-body features but still contain three- and four-body contributions. It also should be noted that many higher-order features can be expressed as products of lower-order features, e.g., $\Tilde{B}^{(4)}_{ijn_{1}n_{2}l_{1}l_{2}0}=\Tilde{B}^{(3)}_{ijn_{1}l_{1}}\Tilde{B}^{(3)}_{ijn_{2}l_{2}}$ and $\Tilde{B}^{(5)}_{ijn_{1}n_{2}n_{3}l_{1}l_{2}l_{3}l_{4}0}=\Tilde{B}^{(4)}_{ijn_{1}n_{2}l_{1}l_{2}l_{4}}\Tilde{B}^{(3)}_{ijn_{3}0}$. In such cases, higher-order features are directly computed as products of lower-order features to reduce the computational cost associated with edge symmetrization for high-order features.

\subsubsection{Output}

The bonding environment of edge $i$-$j$ is represented by a vector $\bm{\zeta}_{ij} = (\zeta_{ij1},\zeta_{ij2},\zeta_{ij3},\ldots)$, whose components are defined as
\begin{align}
\label{eq:zeta}
\zeta_{ijp} =& \sum_{cnl}T_{c\mu_{i}\mu_{j}}W_{cnlp}^{(3)}\Tilde{B}_{ijnl}^{(3)} 
+ \sum_{\substack{cn_{1}n_{2}\\l_{1}l_{2}l_{3}}}T_{c\mu_{i}\mu_{j}}W_{\substack{cn_{1}n_{2}\\l_{1}l_{2}l_{3}p}}^{(4)}\Tilde{B}_{\substack{ijn_{1}n_{2}\\l_{1}l_{2}l_{3}}}^{(4)}
+ \sum_{\substack{cn_{1}n_{2}n_{3}\\l_{1}l_{2}l_{3}l_{4}l_{5}}}T_{c\mu_{i}\mu_{j}}W_{\substack{cn_{1}n_{2}n_{3}\\l_{1}l_{2}l_{3}l_{4}l_{5}p}}^{(5)}\Tilde{B}_{\substack{ijn_{1}n_{2}n_{3}\\l_{1}l_{2}l_{3}l_{4}l_{5}}}^{(5)},
\end{align}
\end{widetext}
where $W^{(3)}$, $W^{(4)}$, and $W^{(5)}$ are trainable weights for the three-, four-, and five-body features, respectively, and $T_{c\mu_{i}\mu_{j}}$ encodes the chemical type of edge $i$-$j$, with $c$ denoting the channel index of the element embedding. The indices $p$ and $c$ take values $1,\ldots,p_\mathrm{max}$ and $1,\ldots,c_\mathrm{max}$, respectively. In practice, the weight tensors $W^{(3)}$, $W^{(4)}$, and $W^{(5)}$ are represented using low-rank tensor decompositions to reduce the number of independent trainable parameters, as described in Appendix~\ref{sec:lowrank}. In this study, the body order of the bonding-environment descriptors is truncated at either three- or four-body correlations to save the computational cost.

The bond order $b_{ij}$ is obtained from the bonding environment descriptor
$\bm{\zeta}_{ij}=(\zeta_{ij1},\zeta_{ij2},\zeta_{ij3},\ldots)$ as 
\begin{equation}
b_{ij}=\Phi(\bm{\zeta}_{ij}),
\end{equation}
where $\Phi$ is a nonlinear function. To describe the asymptotic behavior of bond order $b_{ij}$, $\Phi$ is chosen to satisfy the following conditions: 
\begin{equation}
\label{eq:zero}
\Phi(\bm{0})=1,
\end{equation}
and
\begin{equation}
\label{eq:inf}
\lim_{|\zeta_{ijp}|\rightarrow\infty}
\Phi(\bm{\zeta}_{ij})=0, \qquad \text{for all } p.
\end{equation}
The first condition ensures that the bond energy $V_{ij}$ reduces to the pairwise energy $f(r_{ij})[V_{R}(r_{ij}) - V_{A}(r_{ij})]$ when bond $i$-$j$ has no neighboring atoms. The second condition describes the bond-saturation effect in the Abell-Tersoff bond-order potential. If all neighboring bonds $i$-$k$ ($k\neq i,j$) are indistinguishable from bond $i$-$j$, the magnitude of each component of the bonding-environment descriptor $\bm{\zeta}_{ij}$ increases monotonically with the number of neighboring bonds. Under this assumption, $|\zeta_{ijp}|$ increases without bound as the number of neighboring bonds tends to infinity, and the second condition therefore leads to $b_{ij}\to0$. This limiting condition is expected to improve the extrapolation of BOBOP to highly coordinated or dense systems by suppressing the bond order.

In this work, the following form of $\Phi$ is adopted, which is constructed in two steps.
First, each component of the bond feature vector $\bm{\zeta}_{ij}$ is transformed as
\begin{equation}
\hat{\zeta}_{ijp}=\zeta_{ijp}^{\eta_p},
\label{eq:pow}
\end{equation}
where $\eta_{p}$ are trainable parameters constrained to satisfy $1 \leq \eta_{p} \leq \eta_\mathrm{max}$, with $\eta_\mathrm{max}$ being an adjustable upper bound for numerical stability. This transformation is motivated by Eq.~\eqref{eq:bohatzeta} and allows higher-order correlations to be represented implicitly through products of lower-order features, as described in Eq.~\eqref{eq:zetaexpand}.
In practice, to allow for negative values of $\zeta_{ijp}$, Eq.~\eqref{eq:pow} is replaced by
\begin{equation}
\hat{\zeta}_{ijp}
=
\mathrm{sgn}(\zeta_{ijp})
\left[
(|\zeta_{ijp}|+\epsilon)^{\eta_p}
-\epsilon^{\eta_p}
\right],
\end{equation}
where $\epsilon$ is a small positive constant introduced to ensure numerical stability around $\bm{\zeta}_{ij}=\bm{0}$.

Second, the transformed vector
$\bm{\hat{\zeta}}_{ij}
=(\hat{\zeta}_{ij1},\hat{\zeta}_{ij2},\ldots)$
is mapped to the bond order $b_{ij}$ as
\begin{align}
b_{ij}
=& (1+\|\bm{\hat{\zeta}}_{ij}\|)^{-\sigma}+\sum_{p}v_{p}
\hat{\zeta}_{ijp}
(1+\|\bm{\hat{\zeta}}_{ij}\|)^{-(\sigma+1)}, \label{eq:out} \\
&\|\bm{\hat{\zeta}}_{ij}\| = \sqrt{\sum_{p}\hat{\zeta}_{ijp}^{2}},
\end{align}
where $v_p$ and $\sigma$ are trainable parameters.
The parameter $\sigma$ is constrained to satisfy $\sigma \geq 1$.
The first term in Eq.~\eqref{eq:out} is motivated by Eq.~\eqref{eq:bobij} and satisfies the limiting constraints in Eqs.~\eqref{eq:zero} and~\eqref{eq:inf}. The second term in Eq.~\eqref{eq:out} is empirically introduced to provide additional flexibility in the dependence of the bond order on the bonding-environment descriptor beyond its norm, without retaining the prescribed limiting constraints of the first term.

\subsection{Loss function and optimization}
Parameter optimization of BOBOP is performed by minimizing the loss function
\begin{equation}
\begin{split}
\mathcal{L} =& \rho \frac{1}{N}\sum_{s=1}^{N}\left(\frac{E_{s}-E_{s}^\mathrm{ref}}{n_{s}}\right)^2 \\
&+ \frac{1}{N}\sum_{s=1}^{N}\left[\frac{1}{3n_{s}}\sum_{i=1}^{n_{s}}||\bm{F}_{si}-\bm{F}_{si}^\mathrm{ref}||^2\right],
\end{split}
\label{eq:lossfunc}
\end{equation}
where $N$ is the minibatch size, $s$ indexes a structure in the batch, and $n_{s}$ is the number of atoms in structure $s$.
$E_{s}$ and $E_{s}^\mathrm{ref}$ denote the predicted and reference potential energies, respectively. $\bm{F}_{si}$ and $\bm{F}_{si}^\mathrm{ref}$ are the predicted and reference forces on atom $i$ in structure $s$. The parameter $\rho$ controls the weight of the energy loss relative to the force loss. 
In previous bond-order-based MLIPs \cite{Pun2019,73gp-bm46}, the parameters in the analytical function part were pretrained, whereas in this study all parameters are trained simultaneously starting from randomly initialized values. The isolated-atom energy $\varepsilon(\mu_i)$ was fixed when a reference value was available; otherwise, it was optimized during training.

\section{Asymptotic behavior of bond order}

\label{sec:asymptotic}

\begin{figure}[tbh]
\centering
\includegraphics[clip,scale=0.36]{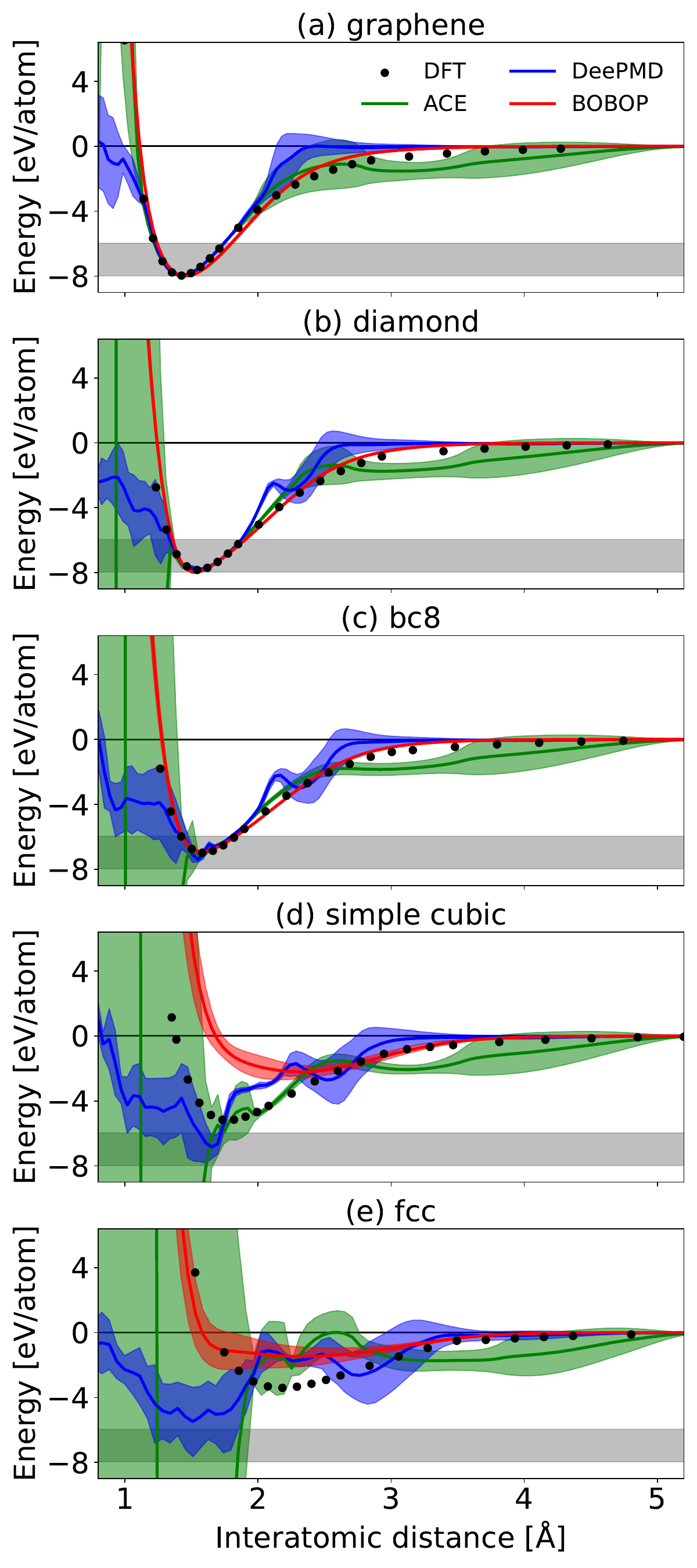}
\caption{Cohesive energy as a function of nearest-neighbor interatomic distances for (a) graphene, (b) diamond, (c) bc8, (d) simple cubic, and (e) fcc, calculated with BOBOP, DeePMD, and ACE trained on the limited-diversity carbon dataset. Each energy curve was averaged over four independently trained models, with the color-shaded regions indicating the standard deviations. The gray-shaded regions indicate the per-atom energy range covered by the training dataset.}
\label{fig:latticelimited}
\end{figure}

\begin{figure}[tbh]
\centering
\includegraphics[clip,scale=0.35]{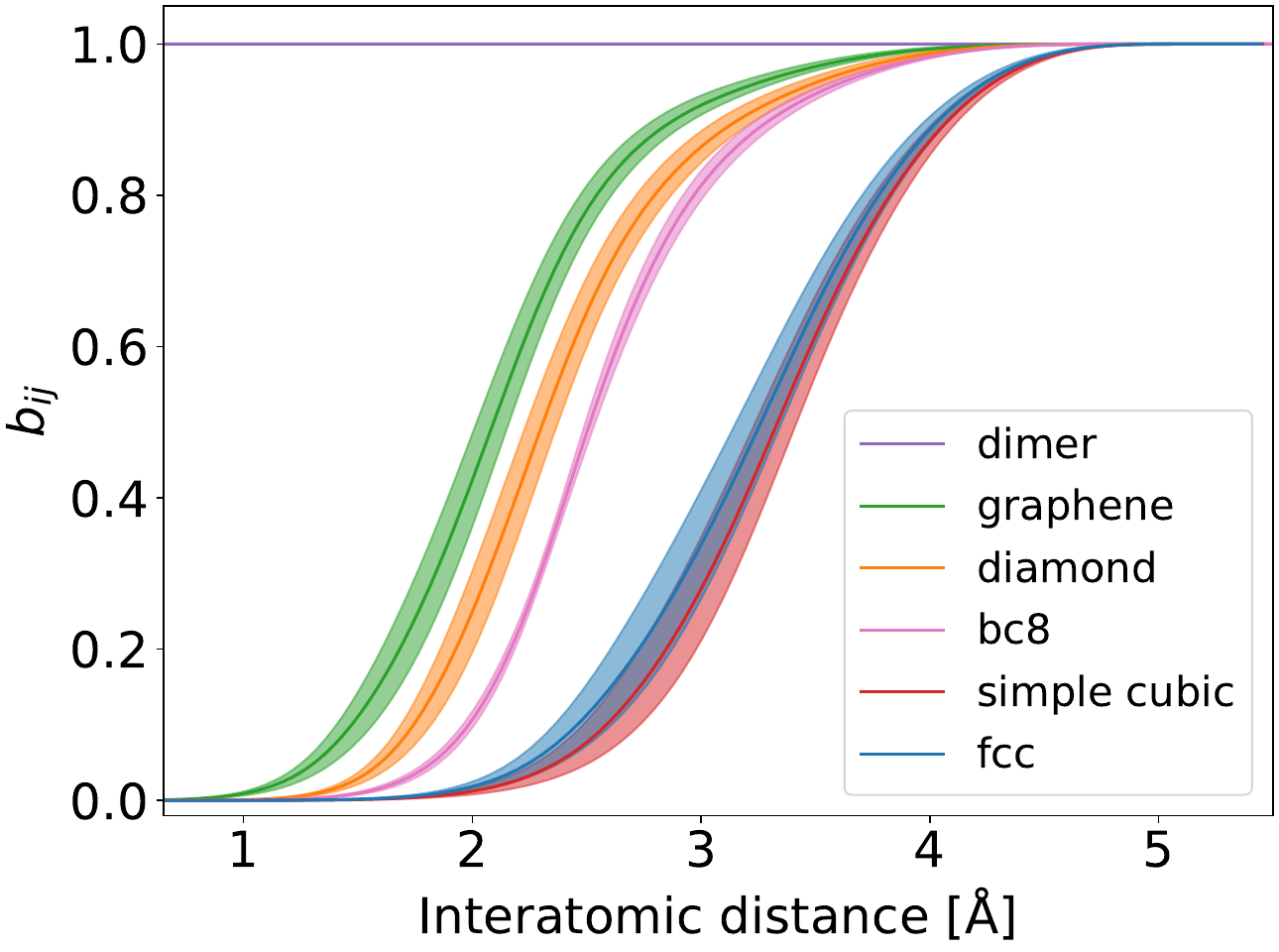}
\caption{Asymptotic behavior of bond order $b_{ij}$ for a nearest-neighbor bond in dimer, graphene, diamond, bc8, simple cubic, and fcc structures under isotropic scaling of the system. The bond-order profiles were averaged over four independently trained models, with the shaded regions indicating half of the standard deviation.}
\label{fig:bondorder}
\end{figure}

To assess the qualitative extrapolation behavior in a simple setting, BOBOP was trained on a limited-diversity carbon dataset comprising 302 configurations of graphene and diamond with randomly perturbed lattice constants and atomic positions. The dataset spans energies up to 2.0 eV/atom above the relaxed graphene structure and includes an isolated atom as a reference for cohesive energies. To assess the sensitivity of the model to stochastic effects during training, four independent training runs were performed using different random seeds. For comparison, two representative local MLIPs, DeePMD \cite{PhysRevLett.120.143001} and ACE, were trained on the same dataset using \textsc{DeePMD-kit} \cite{WANG2018178} and \textsc{pacemaker} \cite{bochkarev2022efficient}, respectively. ACE was trained with L1 regularization using a regularization loss weight of $L_{1} = 10^{-5}$. 

Figure \ref{fig:latticelimited} shows the cohesive energy curves of graphene, diamond, bc8, simple cubic, and fcc predicted by BOBOP, DeePMD, and ACE trained on the limited-diversity dataset. For graphene, diamond, and bc8, BOBOP shows better qualitative agreement with the DFT energy curves in extrapolation regions than the other models. Furthermore, its predictions are highly consistent across four independent training runs, exhibiting only small standard deviations over a wide range of interatomic distances. Although BOBOP does not accurately predict the equilibrium energy minima of the simple cubic and fcc structures, which are absent from the training dataset, it maintains relatively stable gradients over the entire range of interatomic distances.

Figure \ref{fig:bondorder} shows the asymptotic behavior of bond order $b_{ij}$ for a nearest-neighbor bond in various structures, showing that it exhibits the expected bond-saturation behavior. As the system becomes denser and the coordination number increases, the bond order approaches zero in the limit of infinite coordination. Conversely, as the system becomes sparser, the bonding environment approaches that of an isolated dimer, and the bond order approaches one. These asymptotic behaviors of the bond order demonstrate that the bond-order function exhibits the physically motivated asymptotic behavior imposed by the limiting constraints.

\section{Effect of explicit high-order correlation}

\label{sec:highorder}

\begin{figure}[tbh]
\centering
\includegraphics[clip,scale=0.38]{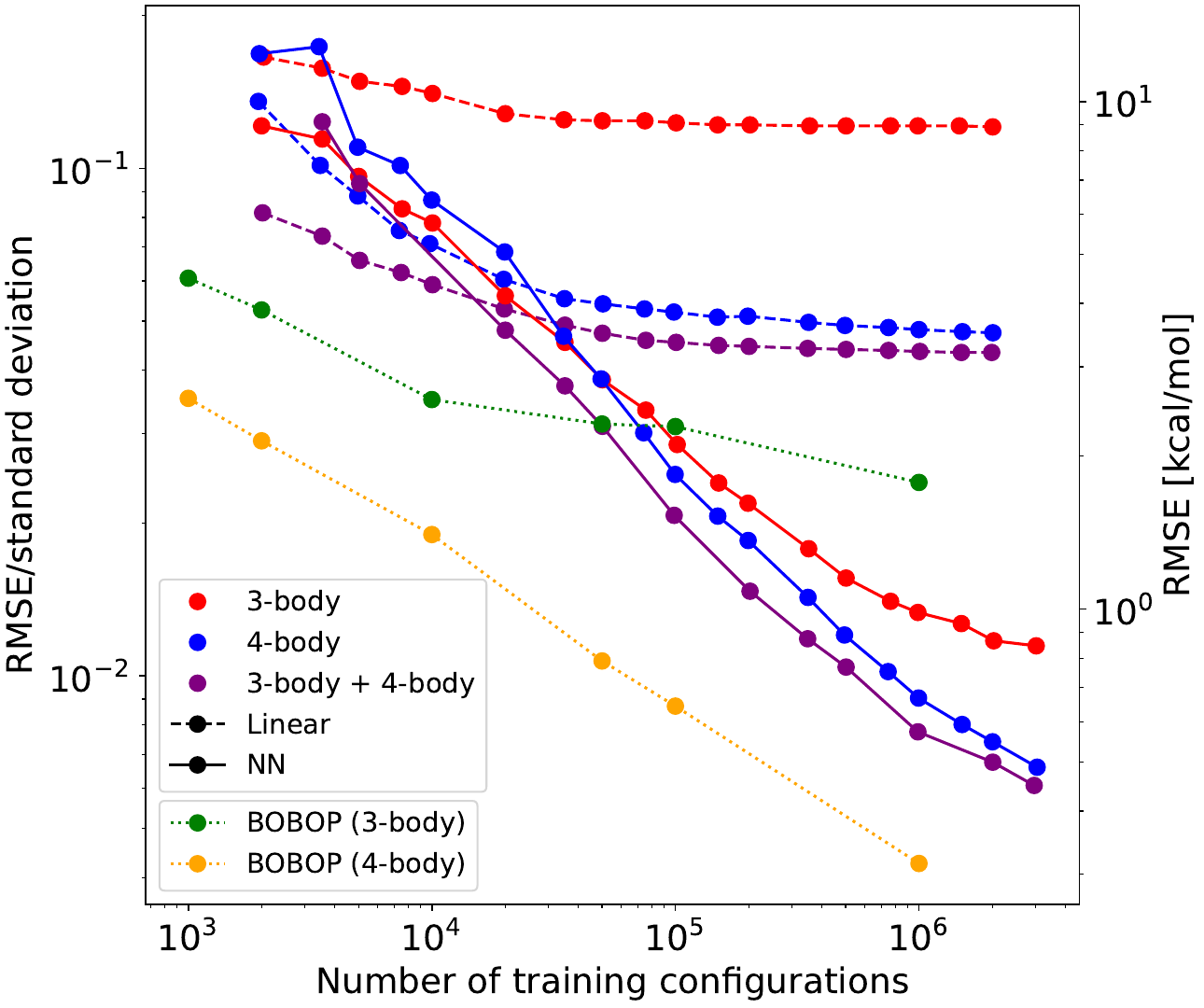}
\caption{Test RMSEs for energy as a function of the number of training configurations of CH$_4$. Results for BOBOP with three- and four-body features are shown. Test RMSEs for three-body, four-body, and combined three- and four-body atom-centered descriptors with linear  and neural-network outputs reported in Ref.~\cite{PhysRevLett.125.166001} are also shown for comparison. The left axis shows the RMSE normalized by the standard deviation of the energy over the CH$_4$ dataset, while the right axis shows the corresponding RMSE in kcal/mol.}
\label{fig:methane}
\end{figure}

\begin{figure}[tbh]
\centering
\includegraphics[clip,scale=0.31]{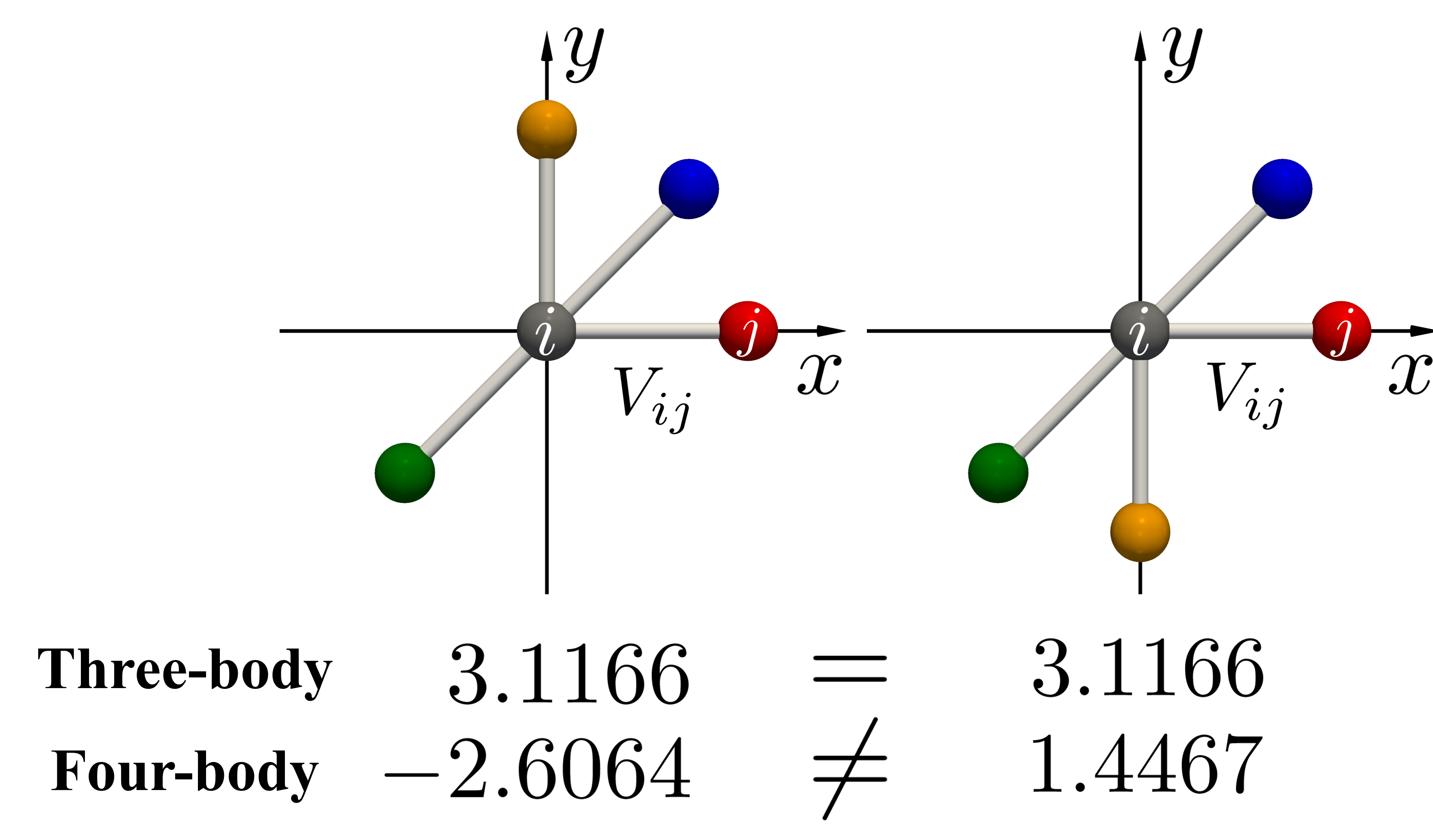}
\caption{Bond-energy degeneracy between two planar CH$_4$ configurations. The bond energy $V_{ij}$ computed with BOBOP based on three-body features exhibits the same value for both configurations, whereas that based on four-body features distinguishes between them.}
\label{fig:bonddegene}
\end{figure}

To investigate the effect of explicit high-order correlations enabled by the body-ordered extension, BOBOP was trained on the CH$_4$ dataset \cite{PhysRevLett.125.166001}, which contains $\sim7.7\times10^{6}$ CH$_4$ configurations with randomly distributed atoms. 80,000 testing configurations were randomly sampled from the CH$_4$ dataset, and the training configurations were sampled from the remaining dataset. 

Figure~\ref{fig:methane} compares the test energy root mean square errors (RMSEs) as a function of the number of training configurations between BOBOP with three- and four-body features and MLIP models based on power spectrum (three-body) and bispectrum (four-body) descriptors combined with linear and neural-network outputs \cite{PhysRevLett.125.166001}. Overall, BOBOP achieves lower test RMSEs with fewer training configurations than the power-spectrum- and bispectrum-based models, suggesting favorable data efficiency. However, the three- and four-body BOBOP models exhibit different learning curves. For the three-body BOBOP, the test RMSE decreases with increasing training set size but eventually reaches a plateau. In contrast, for the four-body BOBOP, the test RMSE continues to decrease as the training set size increases without reaching a clear plateau.

This difference in learning behavior can be understood from the limitation of three-body features in distinguishing distinct bonding environments. As an example, consider two distinct planar CH$4$ configurations, as shown in Fig.~\ref{fig:bonddegene}. Although the two configurations have different arrangements of the neighboring atoms around bond $i$-$j$, the Abell-Tersoff bond-order potential and BOBOP based on three-body features yield the same bond energy $V_{ij}$ because they have identical bond-angle histograms relative to bond $i$--$j$, as explained in Sec.~\ref{sec:background}. In contrast, the inclusion of four-body features provides additional information on correlations among multiple neighboring bonds, allowing the two configurations to be distinguished and thereby resolving the bond-energy degeneracy. This ability to distinguish bonding environments contributes to the improved accuracy and data efficiency of BOBOP with four-body features.


\section{Training on publicly avaiable datasets}
\label{sec:training}
To investigate the interpolation accuracy and extrapolation performance, we train BOBOP on various training datasets and test it on configurations outside the training domain, in comparison with existing MLIP models. Hyperparameter settings of BOBOP are summarized in Appendix~\ref{sec:hypara}.

\subsection{Silicon}

\begin{figure}[tb]
\centering
\includegraphics[clip,scale=0.35]{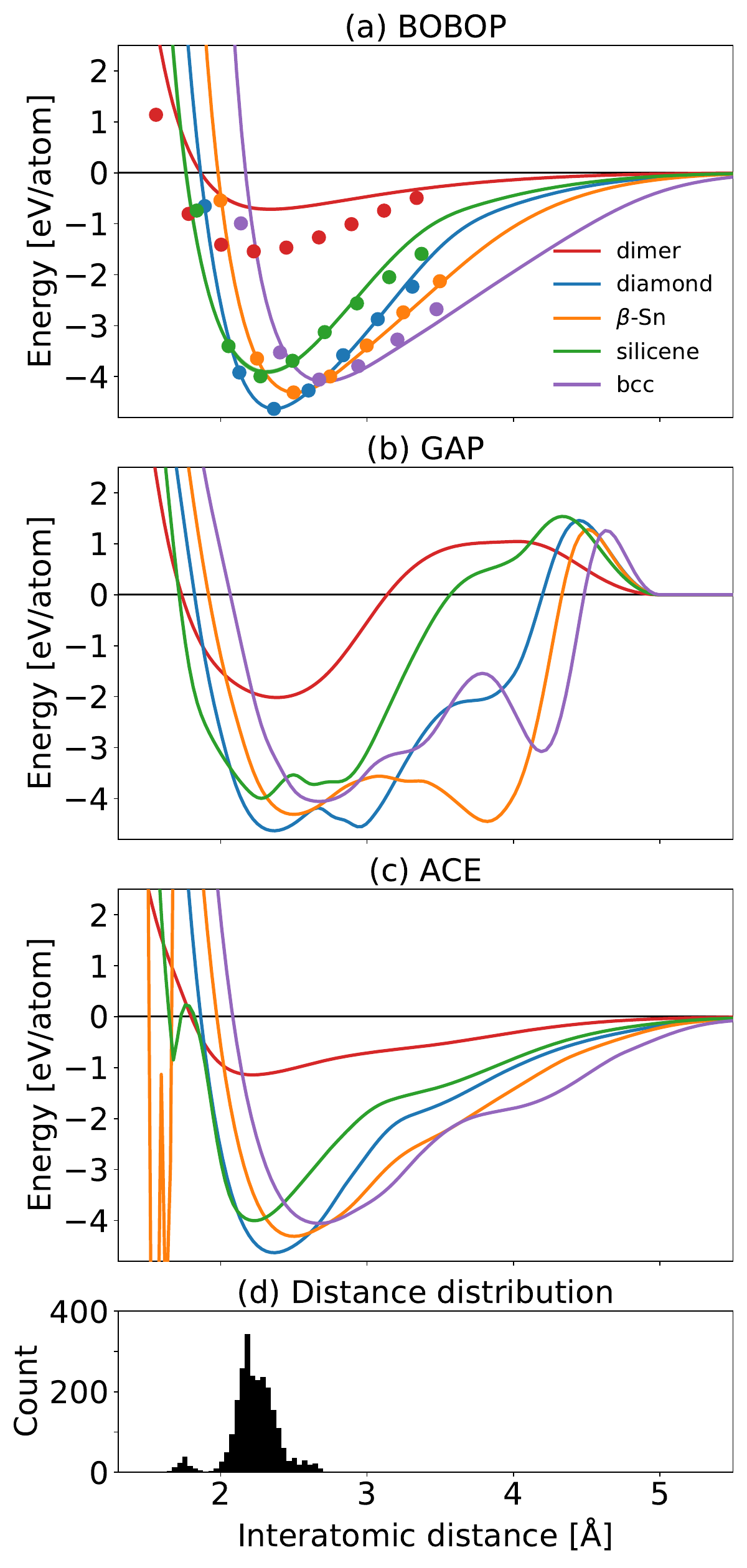}
\caption{Cohesive energy as a function of nearest-neighbor interatomic distances for dimer, diamond, $\beta$-Sn, silicene, and bcc, calculated with (a) BOBOP, (b) GAP, and (c) ACE trained on the GAP silicon dataset. The dots indicate the DFT-calculated energies for each structure. Panel (d) shows the distribution of nearest-neighbor interatomic distances included in the GAP silicon dataset. The BOBOP potential energy curves are averaged over four independently trained models, with shaded regions indicating the standard deviation. }
\label{fig:curveongapsi}
\end{figure}

\begin{figure}[tbh]
\centering
\includegraphics[clip,scale=0.35]{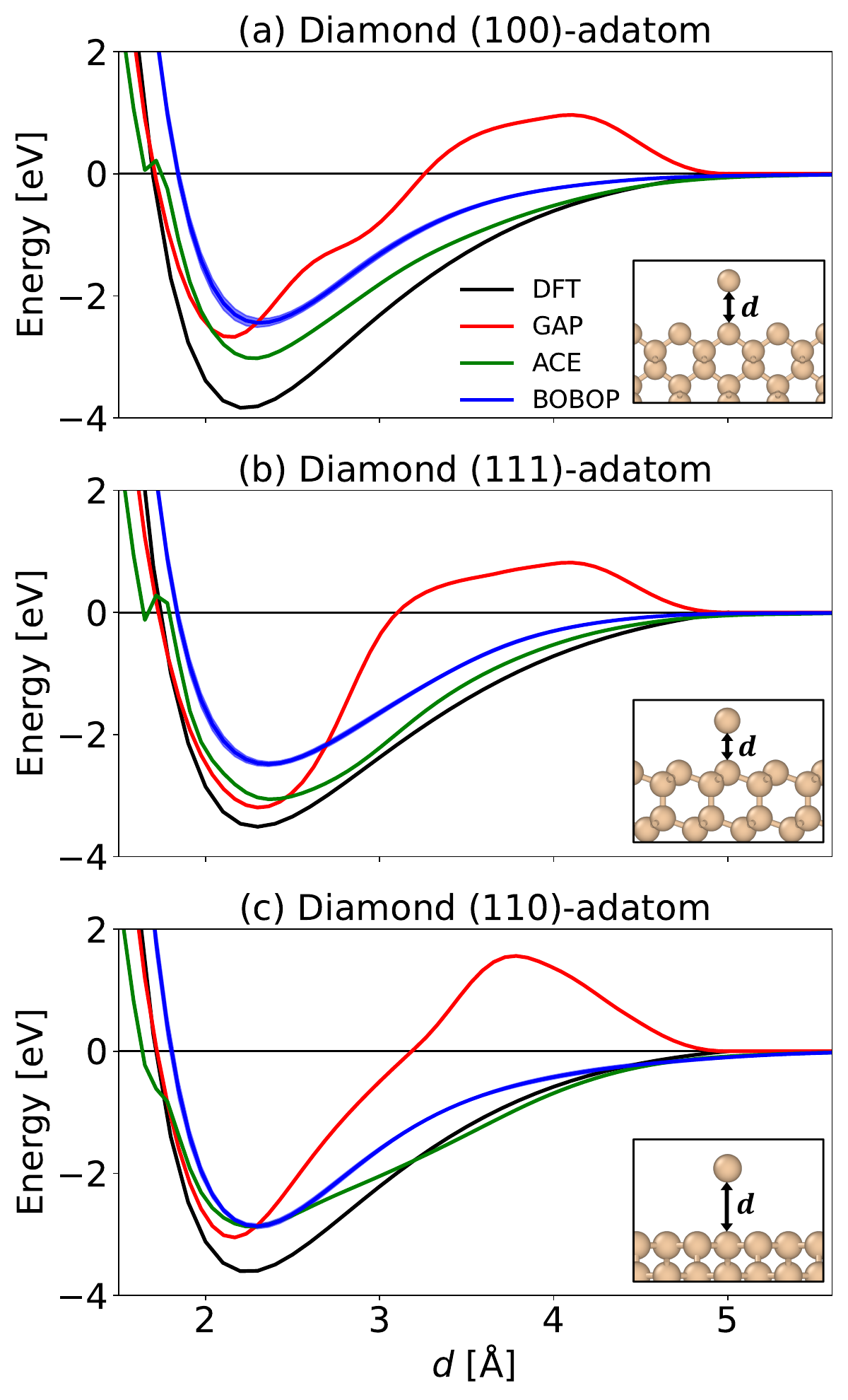}
\caption{Decohesion energy curves for (a) Si (100)-adatom, (b) Si (111)-adatom, and (c) Si (110)-adatom separation, computed using DFT, GAP, ACE, and BOBOP. Adatoms are located at the on-top site of each surface. The BOBOP potential energy curves are averaged over four independently trained models, with shaded regions indicating the standard deviation.}
\label{fig:decohesiongapsi}
\end{figure}

\begin{figure}[tbh]
\centering
\includegraphics[clip,scale=0.38]{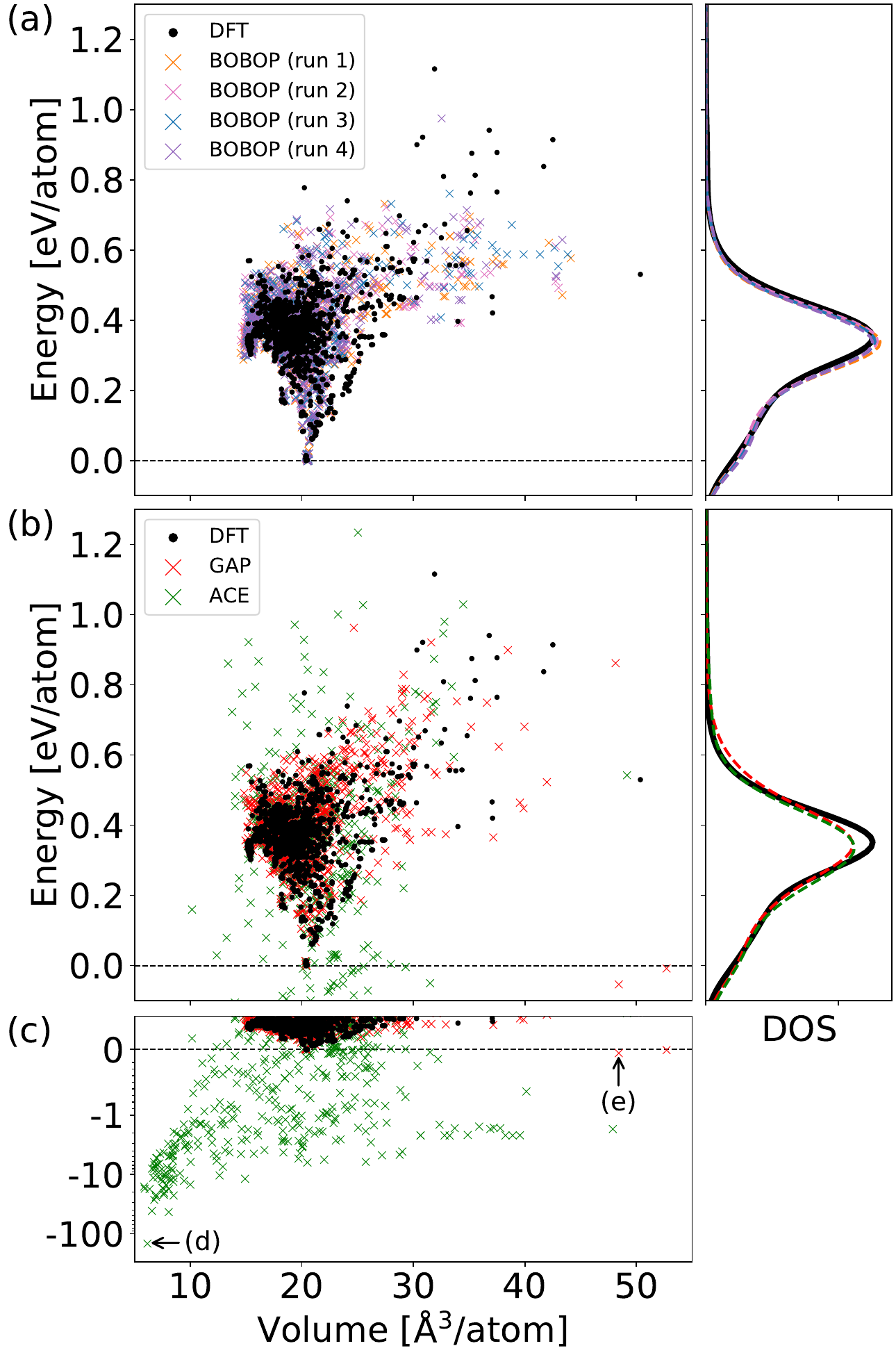}
\includegraphics[clip,scale=0.30]{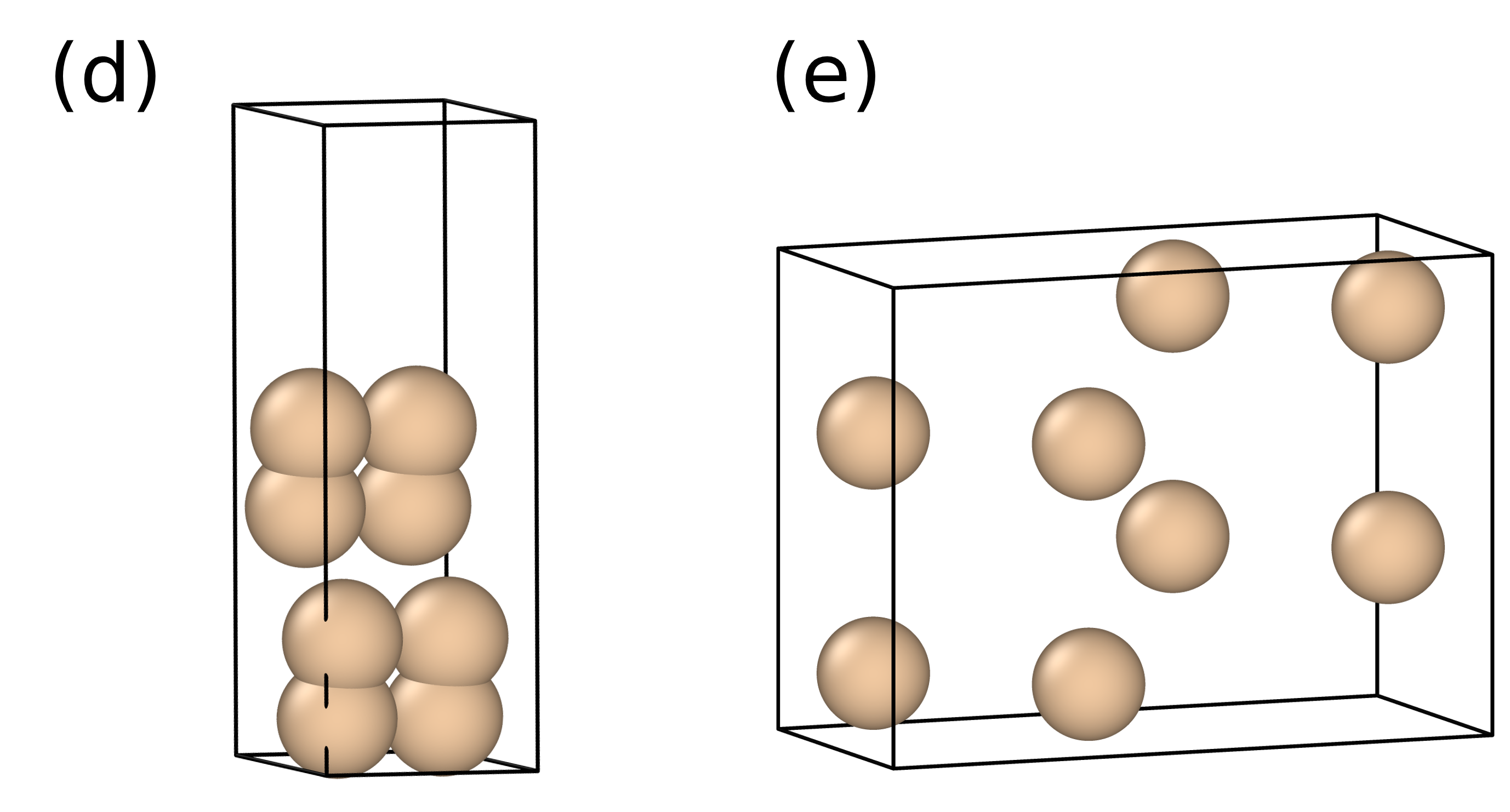}
\caption{Volumes and energies (relative to the diamond structure) of local minima structures obtained with random structure searching (RSS) at 0 GPa. (a) RSS minima distributions obtained using BOBOPs fitted in four independent training runs, in comparison with DFT RSS minima. The right panel shows the corresponding density of states. (b) RSS minima distributions generated using GAP and ACE. (c) Negative-energy region beyond the range of panel (b), with a logarithmic scale on the y-axis. The spurious lowest-energy structures predicted by (d) ACE and (e) GAP are indicated in panel (c) and illustrated in the bottom.}
\label{fig:rsssilicon}
\end{figure}

\begin{figure}[tbh]
\centering
\includegraphics[clip,scale=0.35]{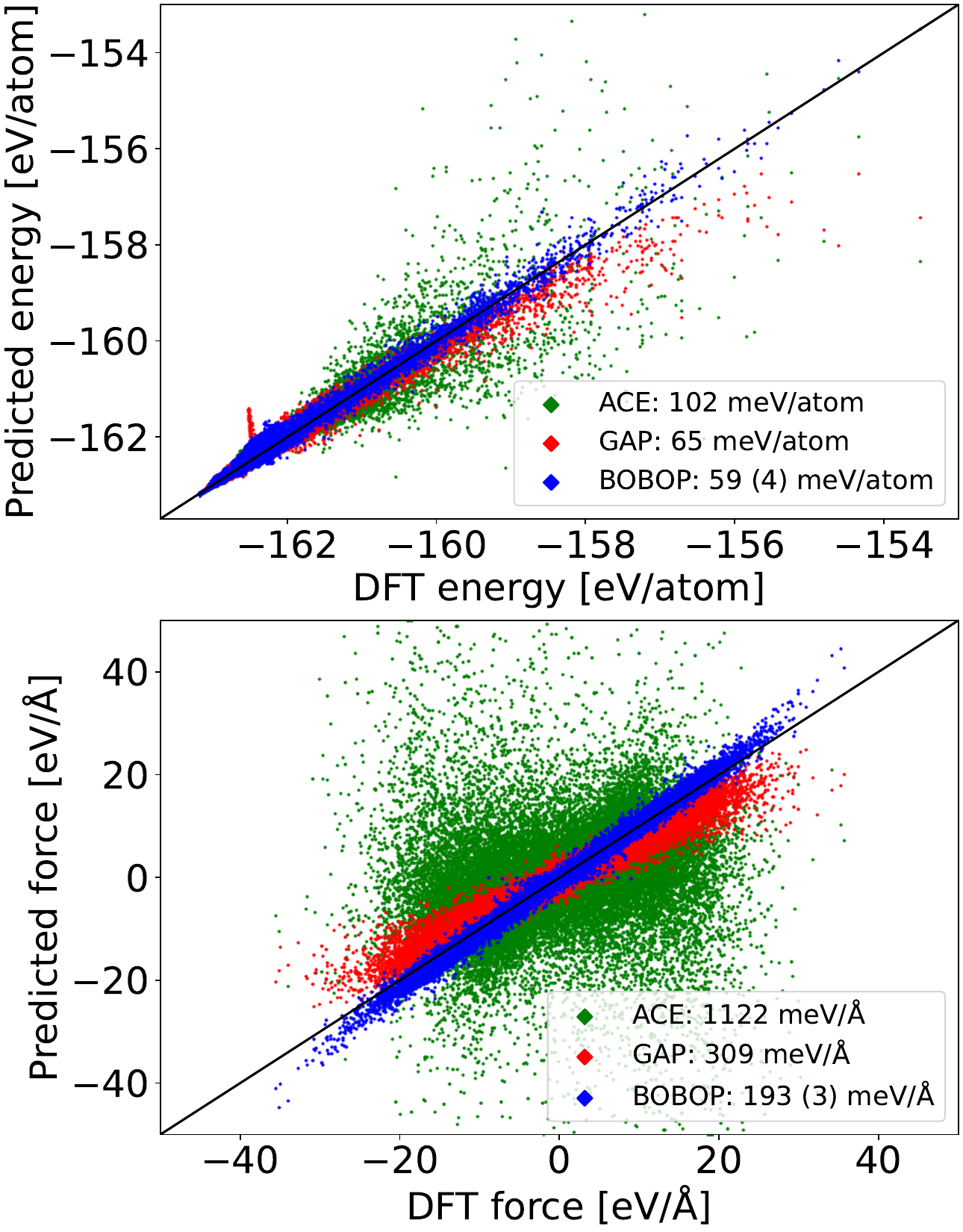}
\caption{Comparison of energy and force predictions by ACE, GAP, and BOBOP for structures along the DFT relaxation paths of RSS at 0 GPa. The corresponding RMSE values are reported in the legends. For BOBOP, the scatter plots show the mean predictions of four independently trained models. The reported RMSE values for BOBOP are averaged over the four models, with the standard deviations given in parentheses.}
\label{fig:rssrmse}
\end{figure}

\begin{table}[tbh]
\caption{\label{table:silicon}
Energy and force RMSEs on the GAP silicon dataset \cite{PhysRevX.8.041048}. For BOBOP, the average RMSEs over four independent training runs are shown, with standard deviations given in parentheses.}
\begin{ruledtabular}
\begin{tabular}{lccc}
 & BOBOP & GAP \cite{PhysRevX.8.041048} & ACE \cite{lysogorskiy2021performant} \\
\midrule
Energy (meV/atom) & 2.59 (0.07) & 1.36 & 3.46 \\
Force (meV/\r{A}) & 76.3 (0.9) & 82.5 & 77.7 \\
\end{tabular}
\end{ruledtabular}
\end{table}

\begin{table}[tbh]
\caption{\label{table:siliconelastic}
Elastic constants (bulk modulus $B$, stiffness tensor components $C_{11}$, $C_{12}$, and $C_{44}$) of diamond-structured silicon, and unreconstructed (but relaxed) surface energies for the (111), (100), and (110) facets. For BOBOP, the average values of four independently trained models are shown, with standard deviations given in parentheses.}
\begin{ruledtabular}
\begin{tabular}{ccccc}
 & BOBOP & GAP \cite{PhysRevX.8.041048} & ACE \cite{lysogorskiy2021performant} & DFT \cite{PhysRevX.8.041048} \\
\midrule
Elastic (GPa) & & & & \\
B  & 81 (2) & 83 & 80 & 82 \\
$C_{11}$ & 145 (3) & 145 & 142 & 147 \\
$C_{12}$ & 49 (1) & 52 & 50 & 50 \\
$C_{44}$ & 76 (2) & 69 & 70 & 73 \\
\midrule 
Surface (J/m$^2$) & & & & \\
(111) & 1.47 (0.03) & 1.50 & 1.47 & 1.56 \\
(100) & 2.08 (0.01) & 2.12 & 2.11 & 2.17 \\
(110) & 1.49 (0.01) & 1.55 & 1.51 & 1.52 \\
\end{tabular}
\end{ruledtabular}
\end{table}

Silicon has been an important target of both MLIPs and bond-order potentials, owing to its complex covalent-bonding potential energy surface. To evaluate the performance of BOBOP, it was trained on the GAP silicon dataset \cite{PhysRevX.8.041048}, which contains 2475 structures with various local environments. For fitting BOBOP, three-body features were used, and four independent training runs were conducted with different random seeds. Table \ref{table:silicon} presents the RMSEs of GAP \cite{PhysRevX.8.041048}, ACE \cite{lysogorskiy2021performant}, and BOBOP on the silicon dataset, showing that BOBOP achieves accuracy comparable to that of the other models. Table \ref{table:siliconelastic} lists the elastic constants and surface energies computed using BOBOP, GAP, ACE, and DFT. The BOBOP results are in good agreement with the DFT values, with an accuracy comparable to that of GAP and ACE.

As an initial validation, the cohesive energy profiles for dimer, diamond, $\beta$-Sn, silicene, and bcc were examined. The energy curves were computed by varying the lattice constants while keeping the fractional coordinates fixed. Figure \ref{fig:curveongapsi} shows the cohesive energy curves as a function of nearest-neighbor interatomic distances computed using DFT, BOBOP, GAP, and ACE, along with the distribution of nearest-neighbor interatomic distances included in the training dataset. All models accurately reproduce the energy curves near equilibrium except for the dimer, which is not included in the GAP silicon dataset. Outside the training domain, however, GAP and ACE exhibit unphysical oscillations in the large and small volume regions, respectively, whereas BOBOP exhibits neither such unphysical minima nor oscillatory behavior and maintains smooth energy profiles over the investigated range.

Figure~\ref{fig:decohesiongapsi} presents the energy curves for the decohesion of an adatom from the on-top site of the (100), (111), and (110) surfaces of diamond-structured silicon. Since these configurations are not included in the GAP silicon dataset, none of the models accurately reproduces the depth of the energy curves. Nevertheless, BOBOP yields smooth and qualitatively correct energy profiles, whereas GAP and ACE exhibit unphysical energy barriers or local minima.

As another validation, random structure searching (RSS) was performed to evaluate the quality of the potential energy surfaces for random configurations. For the RSS, initial configurations were generated by randomly placing eight atoms in randomly shaped cells, with a minimum interatomic spacing of 1.7 \r{A}. The cell volumes were sampled from a uniform distribution in the range of 12--30~\r{A}$^{3}$/atom. A total of 2000 structures were generated, comprising 1000 structures without symmetry constraints and 1000 structures with 2--4 randomly selected symmetry operations. The RSS was first performed using DFT. For cases in which the DFT calculations did not converge, new initial configurations were generated, and the calculations were repeated until a total of 2000 structures were obtained. The RSS was then performed using BOBOP, GAP, and ACE, starting from the same initial configurations used for the DFT RSS. The DFT calculations were performed using the CASTEP package \cite{ClarkSegallPickardHasnipProbertRefsonPayne+2005+567+570} with the same settings and pseudopotentials as those used for the GAP silicon dataset. Structural relaxation was performed using the limited-memory Broyden--Fletcher--Goldfarb--Shanno (L-BFGS) algorithm \cite{Liu1989} with a force tolerance of 0.05~eV/\r{A} for DFT, and the fast inertial relaxation engine (FIRE) algorithm \cite{PhysRevLett.97.170201} with a force tolerance of 0.001~eV/\r{A} for the MLIPs. For BOBOP, the RSS was performed using the four independently trained models.

Figure~\ref{fig:rsssilicon} compares the energy-volume maps and corresponding density of states (DOS) of the RSS minima structures generated by each model. Unlike empirical potentials tested in Ref. \cite{PhysRevX.8.041048}, all models correctly reproduced the position of the DOS peak. Nevertheless, GAP and ACE predicted physically unreasonable structures to be more stable than the diamond structure. Specifically, GAP predicted very sparse minimum structures, with all interatomic distances exceeding 3.6~\r{A}, to have energies lower than that of the diamond structure, whereas ACE predicted highly dense structures containing atoms with coordination numbers exceeding five to be more stable than the diamond structure. In contrast, BOBOP correctly predicted diamond as the lowest-energy structure among the RSS minima for all four independently trained models.

To evaluate the accuracy for structures away from local minima, GAP, ACE, and BOBOP were evaluated on structures generated during the DFT RSS relaxations, comprising 277,536 structures and 2,220,288 local environments in total. Figure~\ref{fig:rssrmse} compares the energy and force predictions of the three models along these DFT RSS relaxation trajectories. Compared with GAP and ACE, BOBOP exhibits better robustness in describing the potential energy surface, with lower errors for configurations far from local minima.

\subsection{Carbon}

\begin{table}[tbh]
\caption{\label{table:mlbop}
Energy and force RMSEs on the ACE carbon dataset \cite{doi:10.1021/acs.jctc.2c01149}. The RMSEs of BOBOP are averaged over four independent training runs, with standard deviation given in parentheses.}
\begin{ruledtabular}
\begin{tabular}{lcc}
 & BOBOP & ACE \cite{doi:10.1021/acs.jctc.2c01149} \\
\midrule
Energy (meV/atom) & 192.7 (20.1) & 166 \\
Force (meV/\r{A}) & 343.4 (2.6) & 689 \\
\end{tabular}
\end{ruledtabular}
\end{table}

\begin{figure}[tbh]
\centering
\includegraphics[clip,scale=0.35]{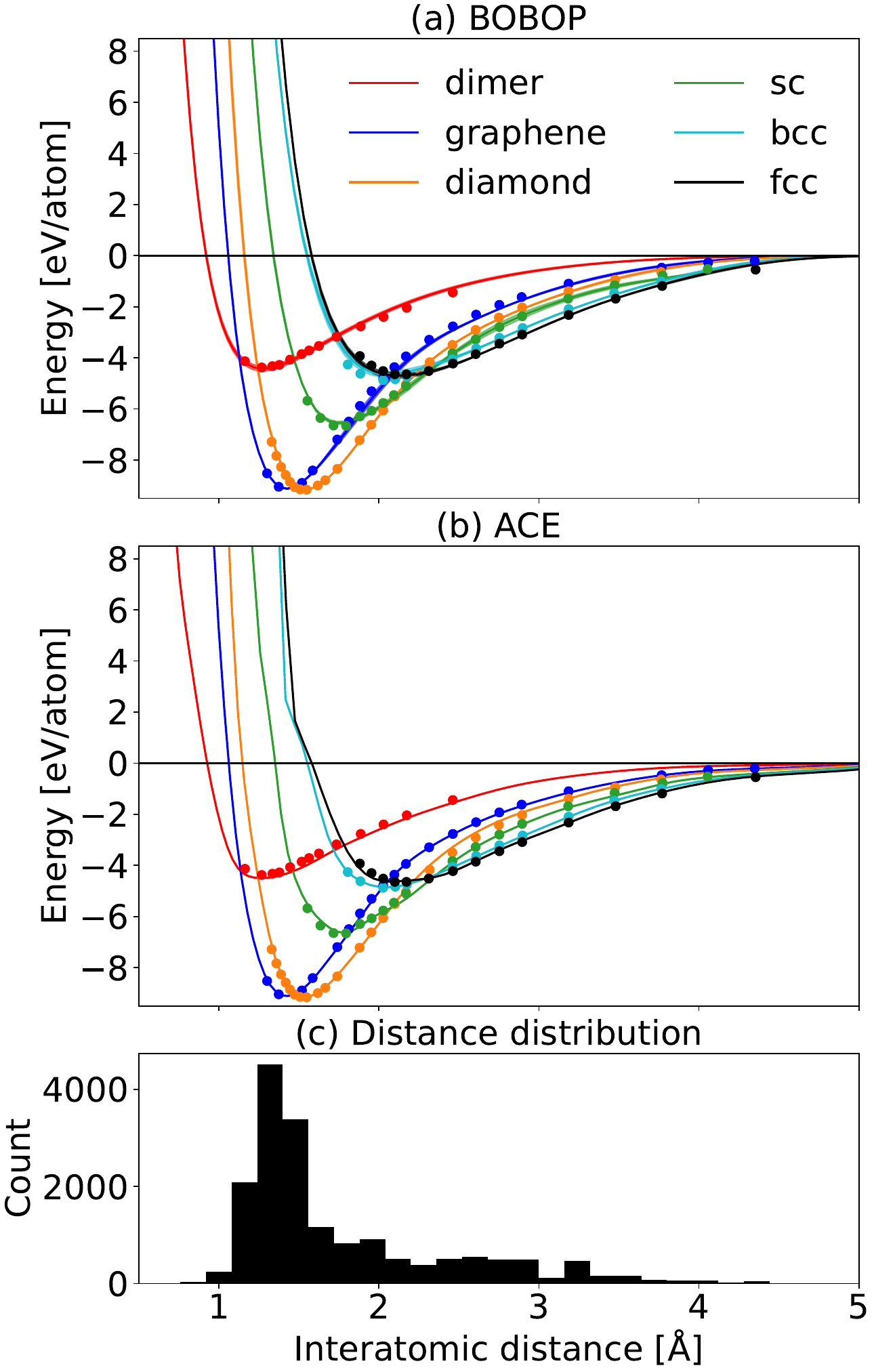}
\caption{Cohesive energy as a function of nearest-neighbor interatomic distances for dimer, graphene, diamond, sc, bcc, and fcc, calculated using (a) BOBOP and (b) ACE with D2 dispersion correction. The dots indicate the cohesive energies obtained with DFT. Panel (c) shows the distribution of nearest-neighbor interatomic distances included in the ACE carbon dataset. The BOBOP energy curves represent the mean values of four independently trained models, with shaded regions showing the standard deviation among the models.}
\label{fig:curveoncarbonace}
\end{figure}

\begin{figure}[tbh]
\centering
\includegraphics[clip,scale=0.36]{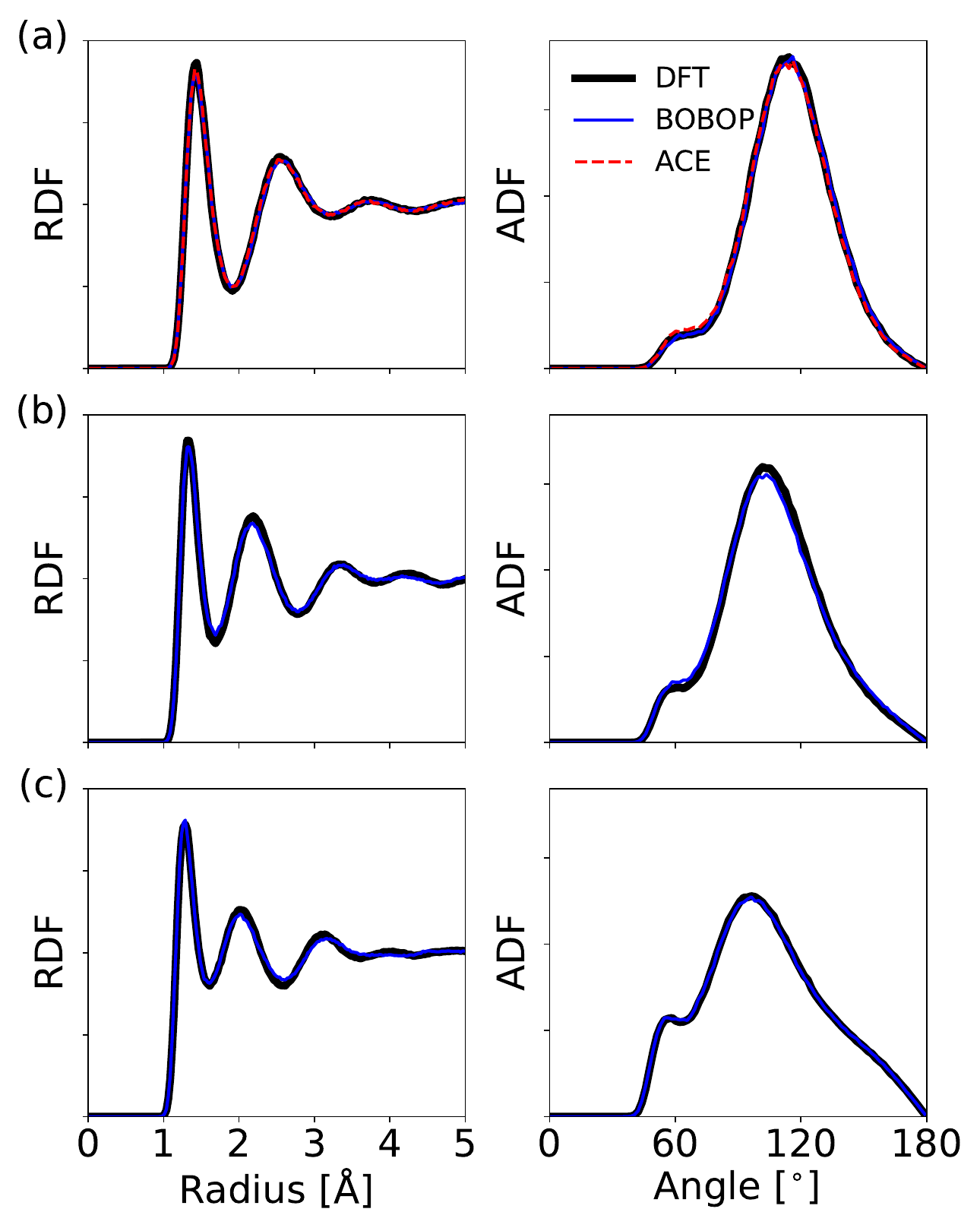}
\caption{RDFs and ADFs obtained from NPT MD simulations of liquid carbon performed using DFT, ACE, and BOBOP at (a) 50 GPa and 7000 K, (b) 500 GPa and 8000 K, and (c) 1000 GPa and 9000 K. The ACE results are not presented in (b) and (c) because the simulations became unstable and the simulation cell collapsed to an unphysically small volume under these conditions. The BOBOP RDFs and ADFs represent the average profiles of four simulations performed using four independently trained models. The cutoff distance for the ADFs was set to 1.6 \r{A}.}
\label{fig:rdfcarbon}
\end{figure}

\begin{figure}[tbh]
\centering
\includegraphics[clip,scale=0.35]{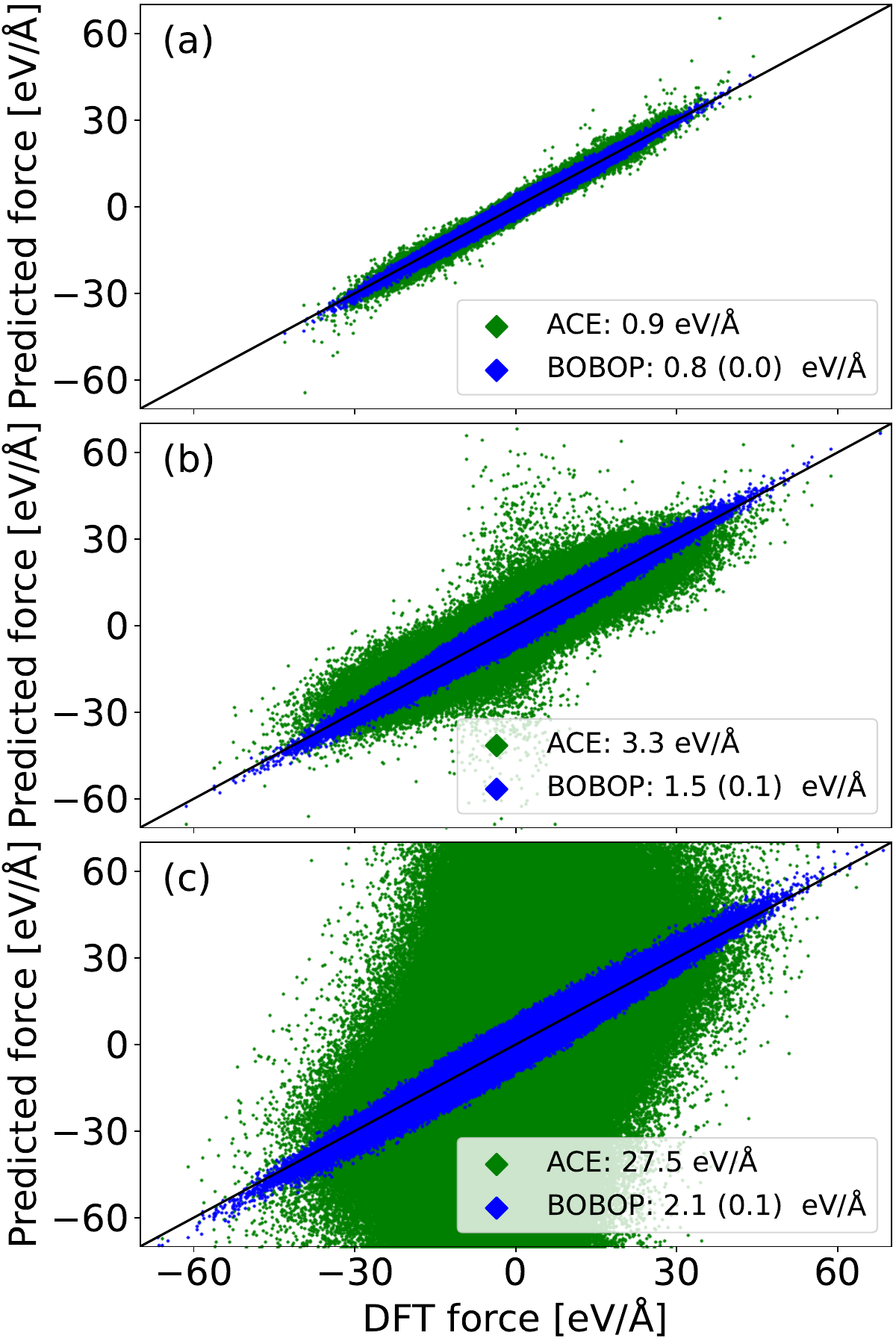}
\caption{Comparison of force predictions by ACE and BOBOP for configurations sampled from DFT MD simulations at (a) 50 GPa and 7000 K, (b) 500 GPa and 8000 K, and (c) 1000 GPa and 9000 K. The corresponding RMSE values are reported in the legends. For BOBOP, the scatter plots show the mean predictions of four independently trained models. The reported RMSE values for BOBOP are averaged over the four models, with the standard deviations given in parentheses.}
\label{fig:rmseforcecarbon}
\end{figure}

\begin{figure}[tbh]
\centering
\includegraphics[clip,scale=0.34]{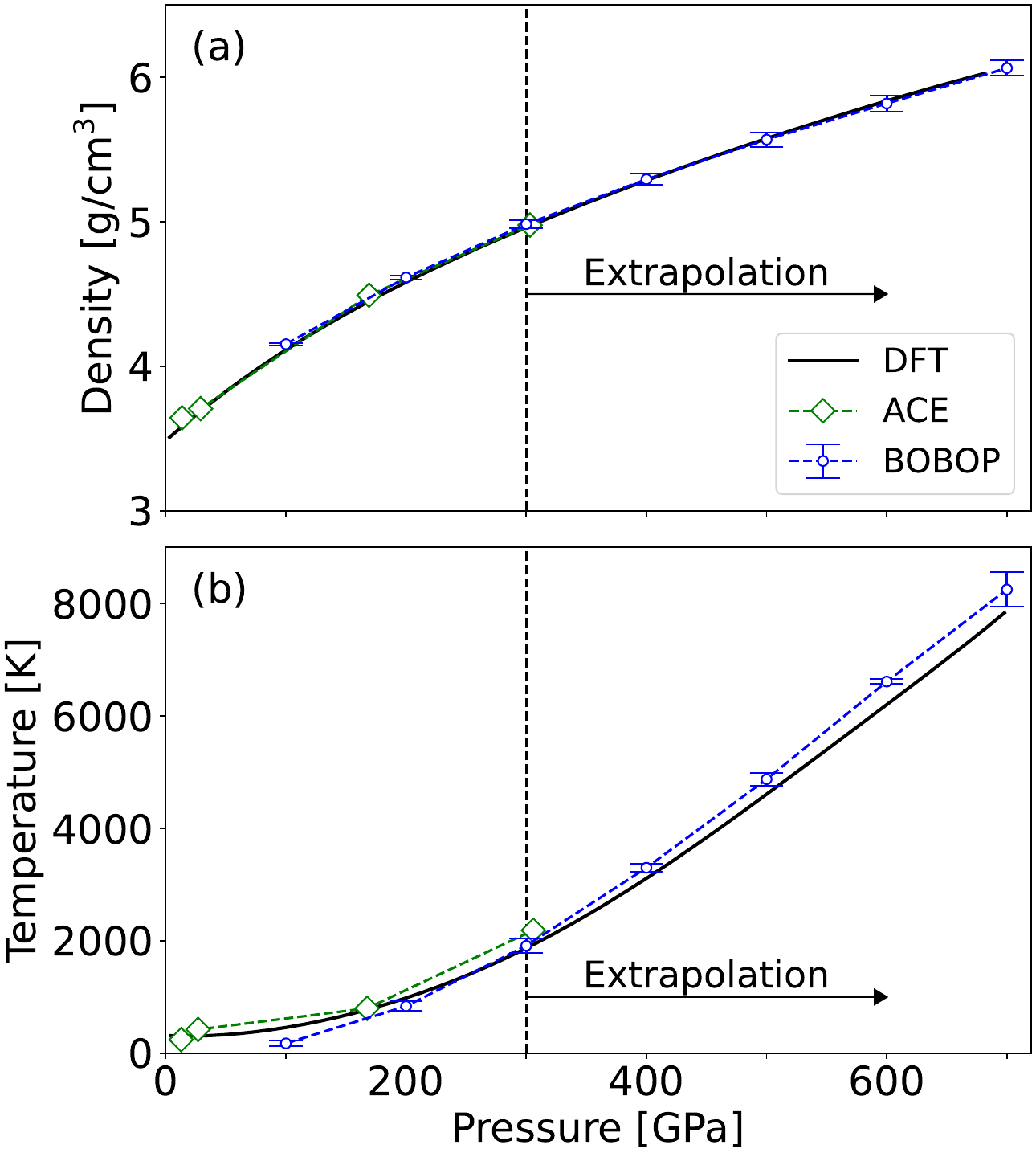}
\caption{Diamond shock Hugoniot curves calculated using DFT, ACE, and BOBOP. (a) Pressure–density Hugoniot curves and (b) pressure–temperature Hugoniot curves. The BOBOP results represent the mean predictions of four independently trained models, with error bars indicating two standard deviations. The DFT and ACE results are from Ref. \cite{10.1063/5.0218705}. }
\label{fig:hugoniotcarbon}
\end{figure}

As another example, BOBOP was trained on the ACE carbon dataset \cite{doi:10.1021/acs.jctc.2c01149}, which includes 17293 structures of carbon with diverse local environments. For training BOBOP, three-body features were used, and four independent training runs were performed with different random seeds. Table~\ref{table:mlbop} presents the RMSEs of BOBOP on the carbon dataset in comparison with ACE, showing that BOBOP achieves accuracy comparable to that of ACE.

Figure \ref{fig:curveoncarbonace} shows the cohesive energy as a function of the nearest-neighbor interatomic distance for the dimer, graphene, diamond, sc, bcc, and fcc structures, calculated using DFT, BOBOP, and ACE. For consistency with Ref. \cite{https://doi.org/10.1002/jcc.20495}, the same D2 correction was added a posteriori to all energy curves shown in Fig. \ref{fig:curveoncarbonace}. The ACE carbon dataset contains a wide range of structures with nearest-neighbor interatomic distances ranging from 0.6 to 5.4 \r{A}. Within this interpolation domain, both BOBOP and ACE accurately reproduce the DFT potential energy surface while maintaining smooth energy profiles.

However, simulations using ACE exhibit unstable behavior at pressures above 300 GPa, as reported in Ref.~\cite{10.1063/5.0218705}, due to the limited coverage of high-pressure configurations in the ACE carbon dataset. To evaluate the extrapolation capability of BOBOP under such high-pressure conditions, isothermal-isobaric (NPT) MD simulations of liquid carbon were performed using DFT, ACE, and BOBOP at 50 GPa and 7000 K, 500 GPa and 8000 K, and 1000 GPa and 9000 K. Each simulation started from a simple cubic structure containing 64 atoms. The DFT MD simulations were performed using VASP \cite{PhysRevB.47.558,Kresse_1994,PhysRevB.54.11169} with a 500 eV cutoff and single $\Gamma$ point sampling, while the MLIP-driven MD simulations were performed using the atomic simulation environment (ASE) \cite{HjorthLarsen_2017}. All simulations were run for 20 ps with a 1 fs timestep. Snapshots collected during the final 10 ps were used for analysis. Figure \ref{fig:rdfcarbon} shows the radial distribution functions (RDFs) and angular distribution functions (ADFs) obtained from DFT, BOBOP, and ACE simulations. At 50 GPa and 7000 K, both BOBOP and ACE reproduce the RDFs and ADFs in good agreement with the DFT results. At pressures above 300 GPa, however, ACE simulations became numerically unstable, consistent with the behavior reported in Ref. \cite{10.1063/5.0218705}. In contrast, BOBOP simulations remain stable and yield RDFs and ADFs in good agreement with DFT even under extreme conditions. 
Figure~\ref{fig:rmseforcecarbon} compares the force predictions of ACE and BOBOP on structures from the DFT MD simulations at each condition. While ACE exhibits increasingly large force deviations under extremely high-pressure and high-temperature conditions, BOBOP maintains accurate force predictions across all investigated conditions. 

As another extrapolation test, BOBOP was evaluated by calculating the diamond shock Hugoniot and comparing the results with DFT and ACE. The shock Hugoniot was obtained from NPT MD simulations of a 216-atom diamond supercell under hydrostatic pressure. The simulations were performed for 30 ps with a timestep of 1 fs, and the shock Hugoniot curves were determined from the time-averaged thermodynamic properties. For each pressure $P$, the Hugoniot temperature $T$ was determined by satisfying the Hugoniot energy conservation equation 
\begin{equation}
E(T,P) - E(T_{0},P_{0}) = \frac{1}{2}(P+P_{0})\left[V(T,P)-V(T_{0},P_{0})\right],
\end{equation}
where $V$ and $E$ are the per-atom internal energy and volume, respectively, and $T_{0}$ and $P_{0}$ are the ambient conditions ($T_{0}=300$ K and $P_{0}=0$ GPa). Figure~\ref{fig:hugoniotcarbon} shows the diamond shock Hugoniot curves obtained using DFT, ACE, and BOBOP. The BOBOP Hugoniot curve agrees well with the DFT results even at high pressures, demonstrating the robustness of BOBOP when extrapolating to high-pressure conditions, where ACE-driven simulations become unstable.

\subsection{Water}

\begin{table*}[tbh]
\caption{\label{table:water}
Test RMSEs of BOBOP on the water dataset \cite{doi:10.1073/pnas.1815117116}, compared with other MLIPs. For non-local models, the number of message-passing layers $T$ is also shown. DeePMD was trained in this work, while other models were trained in previous works. The RMSE of BOBOP (4-body) was averaged over four independent training runs, with standard deviations given in parentheses.}
\begin{ruledtabular}
\begin{tabular}{lccccccccc}
 & BOBOP & BOBOP & \multirow{2}{*}{BPNN \cite{doi:10.1073/pnas.1815117116}} & \multirow{2}{*}{DeePMD}  & \multirow{2}{*}{EANN \cite{D0CP05089J}} & \multirow{2}{*}{ACE \cite{10.1063/5.0158783}} & CACE \cite{Cheng2024} & CACE \cite{Cheng2024} & REANN \cite{PhysRevLett.127.156002} \\
 & (4-body) & (3-body) &  & & & & ($T=0$) & ($T=1$) & ($T=3$) \\
\midrule
Energy (meV/atom) & 1.26 (0.10)& 1.49 & 2.33 & 2.27 & 3.1 & 1.732 & 1.16 & 0.59 & 0.8 \\
Force (meV/\r{A}) & 65.2 (1.3)& 73.6 & 120 & 107 & 129 & 99 & 79 & 47 & 53.2 \\
Cutoff (\r{A}) & 5.5 & 5.5 & 6.2 & 5.5 & 6.2 & 5.5 & 5.5 & 5.5 & 6.2 \\
\end{tabular}
\end{ruledtabular}
\end{table*}

\begin{figure}[tbh]
\centering
\includegraphics[clip,scale=0.38]{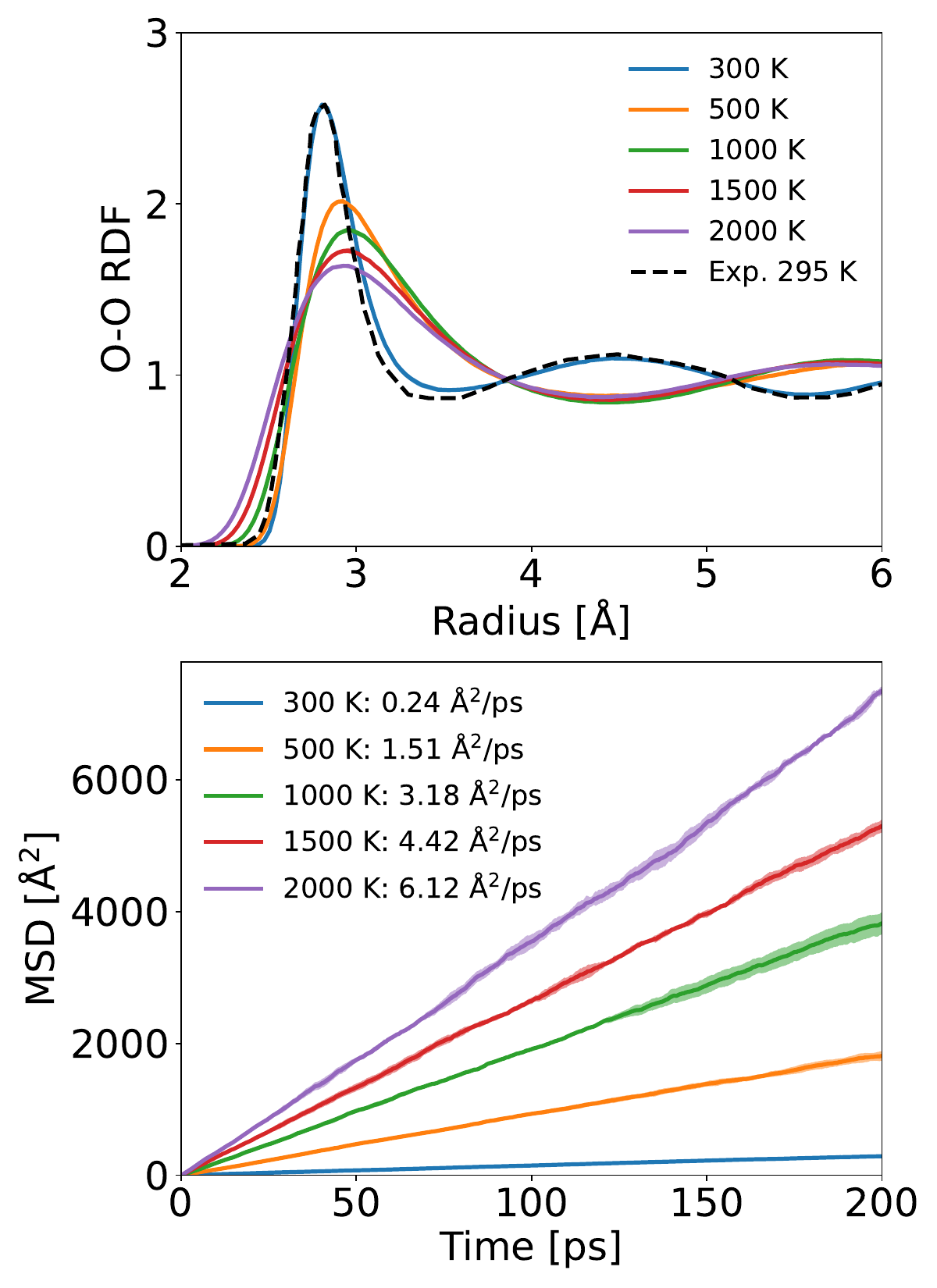}
\caption{Top: Radial distribution functions (RDFs) of water at various temperatures and a density of 1.0 g/cm$^3$, obtained from NVT MD simulations performed using BOBOP (4-body). The RDF from X-ray diffraction \cite{10.1063/1.4902412} is also shown. Bottom: Time history of mean square displacement (MSD) at various temperatures. The BOBOP RDFs and MSDs are averaged over four independently trained models, with shaded regions indicating the standard deviation. The corresponding averaged diffusion constants (\r{A}$^2$/ps) are shown in the legend.}
\label{fig:rdfwater}
\end{figure}

\begin{figure}[tbh]
\centering
\includegraphics[clip,scale=0.32]{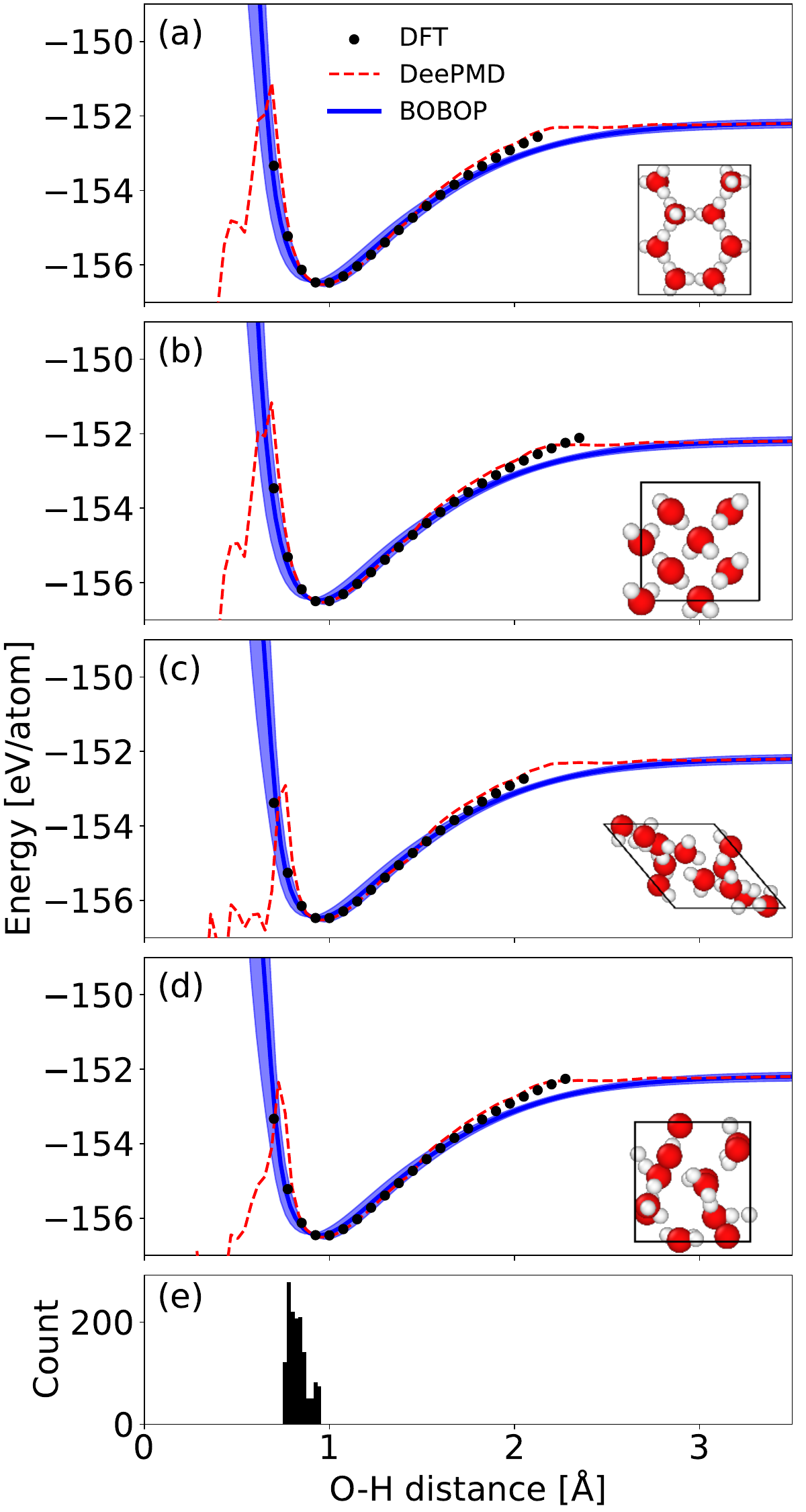}
\caption{Potential energy as a function of nearest-neighbor O-H distance for stable solid phases of water; (a) Ih, (b) Ic, (c) II, and (d) III, computed using DFT, BOBOP, and DeePMD. Energies are relative to the relaxed H$_2$O molecule. Panel (e) shows the distribution of nearest-neighbor O-H distances in the training dataset. The BOBOP potential energy curves are averaged over four independently trained models, with shaded regions indicating the standard deviation.}
\label{fig:latticewater}
\end{figure}

As another application, BOBOP was trained on a water dataset \cite{doi:10.1073/pnas.1815117116} containing 1593 liquid configurations. The dataset was split into 90\% for training and 10\% for testing, as done in Ref. \cite{Cheng2024}. The cutoff distance was set to 5.5~\r{A}, and both three- and four-body features were employed. For BOBOP with four-body features, four independent training runs were performed using different random seeds. Table \ref{table:water} lists the RMSEs on the water dataset of various models; BPNN \cite{doi:10.1073/pnas.1815117116}, EANN \cite{D0CP05089J}, linear ACE \cite{10.1063/5.0158783}, REANN \cite{PhysRevLett.127.156002}, CACE ($T=0,1$) \cite{Cheng2024}, and BOBOP with three- and four-body features. BOBOP with four-body features achieves RMSEs comparable to or lower than those of other local MLIPs: BPNN, EANN, ACE, and CACE ($T=0$).
In contrast, the message-passing MLIPs, REANN and CACE ($T=1$), achieve lower RMSEs, suggesting that accounting for nonlocal information is important for accurately describing systems with spatially heterogeneous charge distributions. However, such nonlocal effects are beyond the scope of the present BOBOP formulation.

As a further validation, MD simulations of liquid water were performed using BOBOP. Figure~\ref{fig:rdfwater} shows the RDFs and the time evolution of the mean square displacement (MSD) at different temperatures and a density of 1.0~g/cm$^3$, obtained from MD simulations of 512 water molecules. The RDFs at all temperatures exhibit physically reasonable features without unphysical minima, indicating that BOBOP maintains a stable potential energy surface over the investigated conditions. At 300 K, the RDFs show good agreement with the experimental results obtained from X-ray diffraction \cite{10.1063/1.4902412}. The MSDs provide a direct measure of the diffusive behavior predicted by BOBOP. At 300 K, BOBOP predicts a diffusion coefficient of 0.240 $\pm$ 0.01~\r{A}$^2$/ps, which is in good agreement with the experimental value of 0.241 $\pm$ 0.015~\r{A}$^2$/ps \cite{B005319H} and substantially closer to experiment than the ACE prediction of 0.120 $\pm$ 0.003~\r{A}$^2$/ps \cite{10.1063/5.0158783}. This agreement suggests that BOBOP provides a physically reasonable description of the dynamical energy landscape.

The potential energy curves of stable solid phases of water, ice Ih, Ic, II, and III, were next examined. The ice structures were generated using the GenIce code \cite{https://doi.org/10.1002/jcc.25077,doi:10.1021/acs.jcim.1c00440}. Figure \ref{fig:latticewater} shows the resulting potential energy curves calculated with BOBOP and DeePMD, along with the distribution of nearest-neighbor O-H distances in the training dataset. The curves were obtained by uniformly scaling the lattice parameters of each ice structure while keeping the fractional coordinates fixed. BOBOP exhibits smooth potential energy curves across the entire range of interatomic distances and maintains physically reasonable asymptotic behavior upon both compression and expansion.

\subsection{Ethanol}

\begin{table}[tbh]
\caption{\label{table:ethanol}
Test MAEs of BOBOP for energy (meV) and force (meV/\r{A}) on ethanol structures from the revMD17 dataset, compared with the other local MLIP models evaluated in Ref.~\cite{doi:10.1021/acs.jctc.1c00647}.  The MAEs of BOBOP are averaged over four independent training runs, with standard deviation given in parentheses.}
\begin{ruledtabular}
\begin{tabular}{lcccccc}
 & BOBOP & ACE & sGDML & FCHL & GAP & ANI\\
\midrule
Energy & 0.87 (0.03) & 1.2 & 2.4 & 0.9 & 3.5 & 2.5 \\
Force & 5.46 (0.18) & 7.3 & 16.0 & 6.2 & 18.1 & 13.4 \\
\end{tabular}
\end{ruledtabular}
\end{table}

\begin{figure}[tbh]
\centering
\includegraphics[clip,scale=0.40]{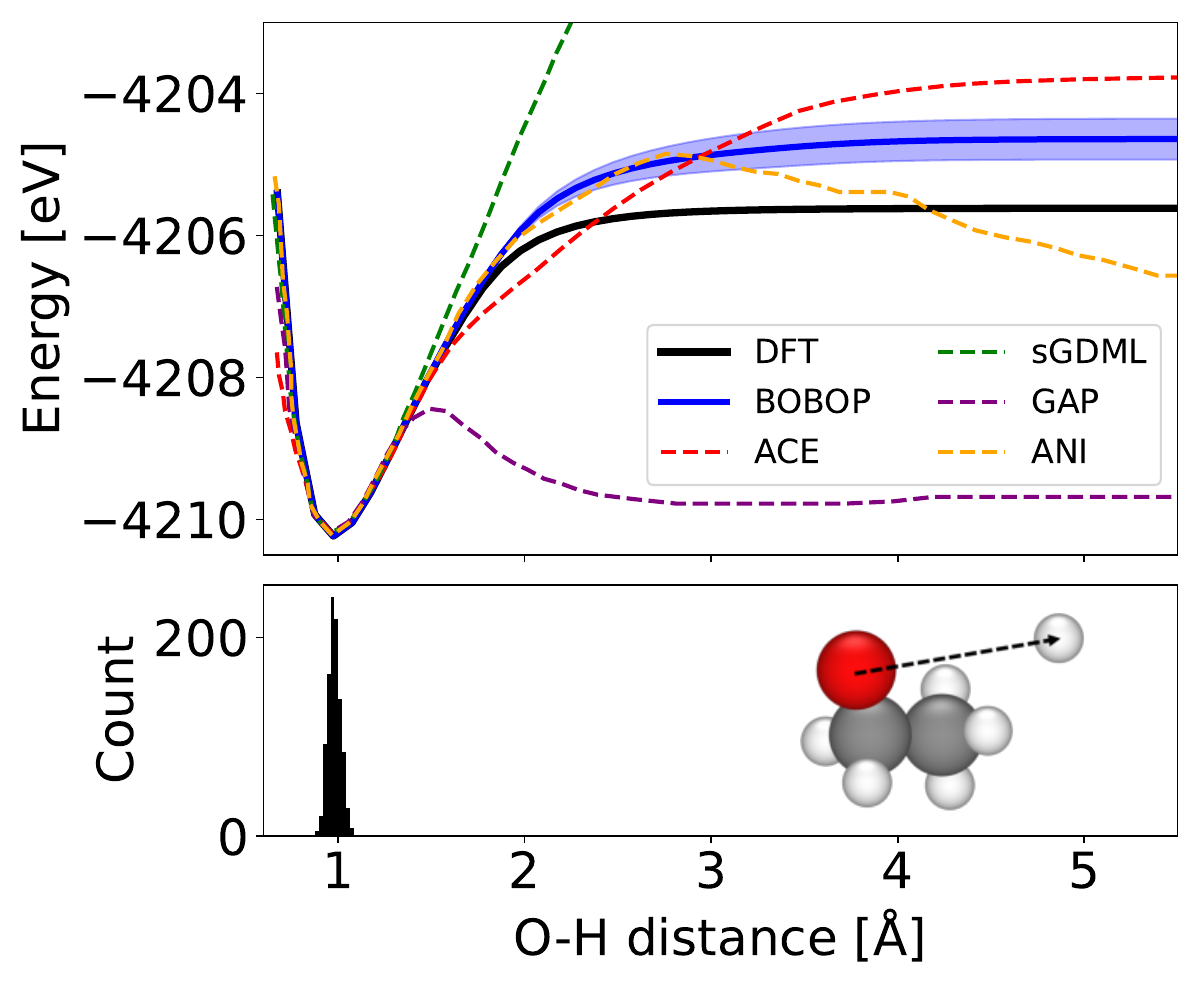}
\caption{Potential energy curves for H removal from ethanol as a function of the O–H distance, calculated using BOBOP, ACE, sGDML, GAP, and ANI trained on the ethanol subsets from the revMD17 dataset. The energy curves for DFT, ACE, sGDML, GAP, and ANI are regenerated from Ref.~\cite{doi:10.1021/acs.jctc.1c00647}. The BOBOP energy curve shows the average prediction of four independently trained models, and the blue shaded region represents the corresponding standard deviation. The bottom panel shows the distribution of O–H distances in the training dataset.}
\label{fig:hremoval}
\end{figure}

To investigate the applicability of BOBOP to small molecules, BOBOP was trained on the ethanol subset in the revised MD17 (revMD17) dataset \cite{Christensen_2020}. The cutoff distance was set to 5.5~\r{A}, four-body features were employed, and four independent training runs were performed using different random seeds. The model was trained on 1000 ethanol structures randomly sampled from the revMD17 dataset and tested on another 1000 structures. Table~\ref{table:ethanol} compares the test MAEs of BOBOP, ACE, sGDML \cite{CHMIELA201938}, FCHL \cite{10.1063/1.5126701}, GAP, and ANI \cite{C6SC05720A} on the ethanol structures in the revMD17 dataset, showing that BOBOP achieves accuracy comparable to that of other local MLIP models.

Figure \ref{fig:hremoval} shows the energy profiles for H-atom removal from the ethanol molecule as a function of O-H distances, computed with BOBOP, DFT, and other MLIP models. The ACE result corresponds to the ACE E0 REG model reported in Ref.~\cite{doi:10.1021/acs.jctc.1c00647}, which uses the isolated atom energy as the one-body term and was trained with regularization. All models reproduce the energy profile reasonably well near equilibrium but exhibit some deviations from DFT at large O--H distances. Accurate prediction in the large distance limit is challenging because the ethoxy radical ($\mathrm{CH_3CH_2O-}$) is absent from the training dataset. Nevertheless, BOBOP maintains a relatively smooth energy profile that is in qualitative agreement with the DFT profile compared to the other models.

\subsection{3BPA}

\begin{table*}[tbh]
\caption{\label{table:3bpa}
Test RMSEs for energy (meV) and force (meV/\r{A}) on the 3BPA datasets, in comparison with local MLIP models tested in Ref. \cite{doi:10.1021/acs.jctc.1c00647}. The models, except for ANI and ANI-2x, were fitted to the 3BPA training dataset at 300 K. ANI-2x was trained on 8.9 million structures \cite{10.1021/acs.jctc.0c00121}. ANI was initialized with ANI-2x weights and finetuned to the 3BPA training dataset at 300 K. The RMSEs of BOBOP are averaged over four independent training runs, with standard deviation given in parentheses.}
\begin{ruledtabular}
\begin{tabular}{lccccccccc}
 & BOBOP & ACE \cite{doi:10.1021/acs.jctc.1c00647} & sGDML \cite{doi:10.1021/acs.jctc.1c00647} & GAP \cite{doi:10.1021/acs.jctc.1c00647} & ANI \cite{doi:10.1021/acs.jctc.1c00647} & ANI-2x \cite{doi:10.1021/acs.jctc.1c00647} & CACE \cite{Cheng2024} & BOTNet \cite{Batatia2025} & MACE \cite{NEURIPS2022_4a36c3c5} \\
\midrule
300 K Energy  & 8.77 (0.13) & 7.1 & 9.1 & 22.8  & 23.5 & 38.6 & 6.3 & 3.1 & 3.0 \\
300 K Force   & 25.0 (0.3) & 27.1 & 46.2 & 87.3 & 42.8 & 84.4 & 21.4 & 11.0 & 8.8 \\
[2pt]
600 K Energy  & 30.0 (4.5) & 24.0 & 484.8 & 61.4  & 37.8 & 54.5 & 18.0 & 11.5 & 9.7 \\
600 K Force   & 52.3 (1.4) & 64.3 & 439.2 & 151.9  & 71.7 & 102.8 & 45.2 & 26.7 & 21.8 \\
[2pt]
1200 K Energy & 69.6 (2.9) & 85.3 & 774.5 & 166.8  & 76.8 & 88.8 & 58.0 & 39.1 & 29.8 \\
1200 K Force  & 118.1 (0.9) & 187.0 & 711.1 & 305.5  & 129.6 & 139.6 & 113.8 & 81.1 & 62.8 \\
\end{tabular}
\end{ruledtabular}
\end{table*}

\begin{figure*}[tb]
\centering
\includegraphics[clip,scale=0.25]{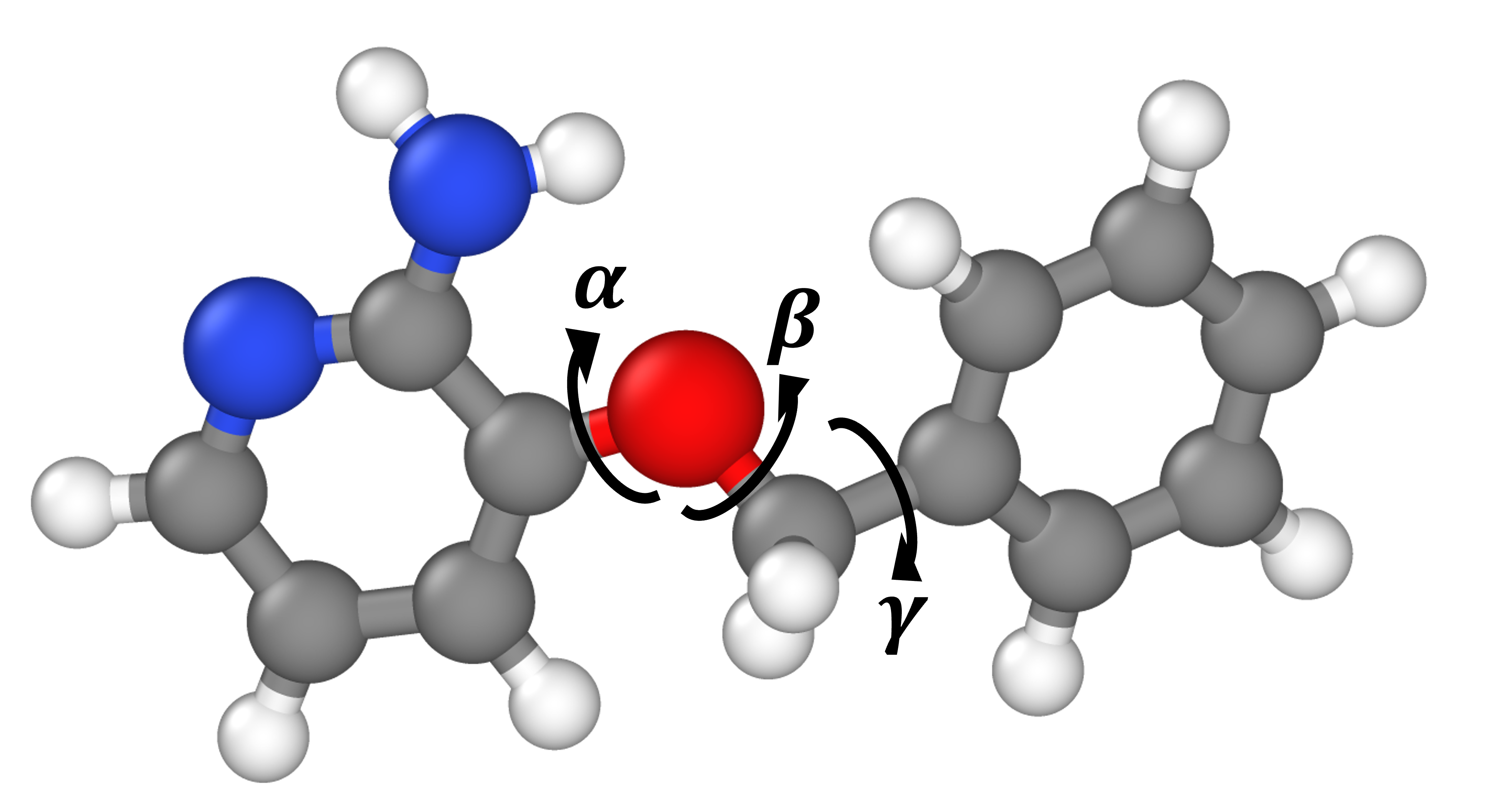}
\includegraphics[clip,scale=0.35]{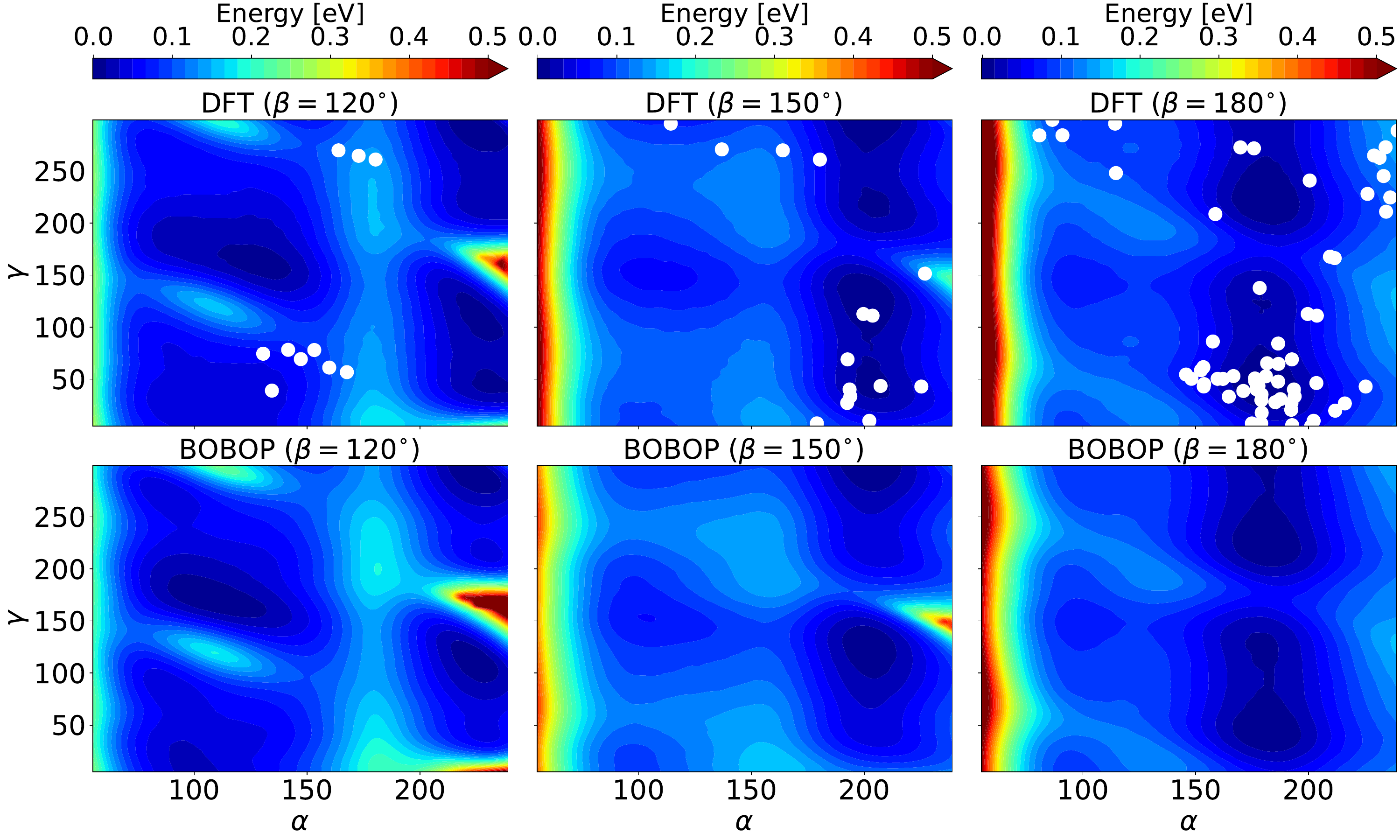}
\caption{Dihedral potential energy landscape of 3BPA at $\beta=$$120^\circ$, $150^\circ$, and $180^\circ$, computed using DFT and BOBOP. The BOBOP landscapes are averaged over four independently trained models. White dots on the DFT landscapes indicate the training dataset configurations that have $\beta$ within $\pm 20^\circ$ of each angle.}
\label{fig:contour}
\end{figure*}

As another example, BOBOP was trained and tested on the 3-(benzyloxy)pyridin-2-amine (3BPA) dataset \cite{doi:10.1021/acs.jctc.1c00647}, which was generated from MD simulations at 300, 600, and 1200 K. BOBOP was trained on the 300 K subset and tested on the 600 and 1200 K subsets. For training BOBOP, four-body features were employed, the cutoff distance was set to 5.5~\r{A}, and four independent training runs were performed.
Table \ref{table:3bpa} compares the RMSEs on the 3BPA test dataset at 300 K, 600 K, and 1200 K for BOBOP, local MLIP models evaluated in Ref.~\cite{doi:10.1021/acs.jctc.1c00647} (ACE, sGDML, GAP, ANI, and ANI-2x), and message-passing MLIPs based on body-ordered descriptors (CACE, BOTNet \cite{Batatia2025}, and MACE \cite{NEURIPS2022_4a36c3c5}). BOBOP exhibits interpolation accuracy at 300 K comparable to ACE, while more accurately predicting the energies and forces of out-of-distribution 3BPA configurations at 1200 K. The message-passing MLIPs, CACE, BOTNet, and MACE, generally achieve higher accuracy than BOBOP and other local MLIPs, likely due to the effective high-order correlations captured by their message-passing architectures.

The 3BPA molecule has three internal rotational degrees of freedom, $\alpha$, $\beta$, and $\gamma$, as illustrated at the top of Fig.~\ref{fig:contour}. To assess the quality of the potential energy landscape, BOBOP was tested on the dihedral potential energy landscape of 3BPA by varying the angles $\alpha$ and $\gamma$ while keeping $\beta$ fixed.
Figure \ref{fig:contour} compares the $\alpha$–$\gamma$ potential energy landscapes at $\beta=120^\circ$, $150^\circ$, and $180^\circ$ between DFT and BOBOP. Overall, BOBOP reproduces the DFT energy landscapes well, even in extrapolation regions far from the training data points. 

It should be noted that several engineering techniques for MLIPs, such as regularization \cite{doi:10.1021/acs.jctc.1c00647} and internal normalization and data normalization \cite{Batatia2025}, have been reported to significantly improve extrapolation performance on the 3BPA dataset. Although these techniques may further improve the performance of BOBOP, we do not investigate their effects here.

\section{Discussion}
\label{sec:discussion}
In Sec.~\ref{sec:training}, the extrapolation performance of BOBOP was demonstrated for various systems. The robust extrapolation capability observed in these tests may be related to several aspects of its functional form.
One factor is the use of a distance-dependent factor with an exponential decay function motivated by the Abell--Tersoff bond-order potential. Compared to Gaussian, polynomial, or Fourier radial basis functions, this type of radial function can suppress unphysical oscillations, thereby providing more physically reasonable behavior outside the training domain. 

A second important factor is the explicit constraint on the asymptotic behavior of the bond-order function $b_{ij}$. In BOBOP, the functional form of $b_{ij}$ is constructed to approach physically motivated limiting values in the limit of large and small magnitudes of the bonding-environment descriptors. Such constraints help prevent unphysical behavior of the bond order, contributing to the robust extrapolation performance of BOBOP.

A third factor may be the effective high-order correlations implicitly incorporated into the bond-order formulation. The nonlinear function applied to the rotationally invariant edge features in the Abell--Tersoff potential and BOBOP implicitly generates higher-order correlations at the atomic-energy level (see Appendix \ref{sec:relationship}). These high-order correlations can contribute to the accuracy and transferability of BOBOP and may partly explain why BOBOP based on only three-body features can achieve performance comparable to that of MLIPs based on atom-centered descriptors with four-body or higher-order correlations. However, a systematic comparison of the effective body order arising from nonlinear transformations of invariant edge features with the body order of atom-centered descriptors remains to be investigated to assess their contributions to accuracy and transferability.

\label{sec:hypara}
\begin{table*}[tbh]
\caption{\label{table:hyperparameter}
Hyperparameter settings of BOBOP. Listed are the body order of bonding environment descriptor, the maximum angular momentum number $l_\mathrm{max}$, the number of radial basis functions $n_\mathrm{max}$, the number of chemical encoding channels $c_\mathrm{max}$, the latent dimension size for the low-rank approximation of the $\nu$-body weight tensors $q_{\nu,\mathrm{max}}$, the size of bonding environment descriptor $p_\mathrm{max}$, the number of trainable parameters (\# parameters), the cutoff distance $R_\mathrm{cut}$.}
\begin{ruledtabular}
\begin{tabular}{lcccccccc}
& body order & $l_\mathrm{max}$ & $n_\mathrm{max}$ & $c_\mathrm{max}$ & $(q_{3,\mathrm{max}},q_{4,\mathrm{max}},q_{5,\mathrm{max}})$ & $p_\mathrm{max}$ & \# parameters & $R_\mathrm{cut}$ [\r{A}] \\ 
\midrule
CH$_4$ (3-body) & 3  & 6 & 12 & 4 & $(48,\--,\--)$ & 1024 & 348510 & 6.5 \\
CH$_4$ (4-body) & 4 & 4 & 6 & 4 & $(84,128,\--)$ & 128 & 615200
 & 6.5 \\
Limited-diversity dataset & 3 & 6 & 6 & 1 & $(6,\--,\--)$ & 256 & 11321
 & 5.5 \\
Silicon & 3 & 6 & 12 & 1 & $(12,\--,\--)$ & 256 & 22193
 & 6.5 \\
Carbon & 3 & 6 & 12 & 1 & $(12,\--,\--)$ & 256 & 22193
 & 5.5 \\
Water (3-body) & 3 & 4 & 6 & 4 & $(24,\--,\--)$ & 256 & 31872 & 5.5 \\
Water (4-body) & 4 & 4 & 6 & 4 & $(24,12,\--)$ & 64 & 35520 & 5.5 \\
Ethanol & 4 & 4 & 6 & 6 & $(36,36,\--)$ & 32 & 53059 & 5.5 \\
3BPA & 4 & 4 & 6 & 8 & $(36,36,\--)$ & 32 & 56219 & 5.5 \\
\end{tabular}
\end{ruledtabular}
\end{table*}

\label{sec:computation}
\begin{table*}[tbh]
\caption{\label{table:timings}
Model comparison of inference time ($\mu$s/atom/step) and cutoff distance (\r{A}).
Timing was performed using a single CPU core of an Intel Xeon Platinum 8480+ 2.0 GHz.}
\begin{ruledtabular}
\begin{tabular}{lcccccc}
& \multicolumn{3}{c}{Silicon} & \multicolumn{3}{c}{Water} \\
\cmidrule{2-4}\cmidrule{5-7}
& BOBOP & ACE & GAP & BOBOP (three-body) & BOBOP (four-body) & DeePMD \\
\midrule
Timing & 901 & 682 & 6993 & 925 & 9314 & 1348 \\
Cutoff
& 6.5 & 5.5/6.5\footnote{For ACE, the cutoff distance is 5.5 \r{A} for the many-body contribution and 6.5 \r{A} for the pairwise contribution.}
& 5.0
& 5.5 & 5.5 & 5.5 \\
\end{tabular}
\end{ruledtabular}
\end{table*}

\section{Conclusion}
\label{sec:conclusion}
\subsection{Summary}
In summary, this study has presented a semiparametric interatomic potential, named BOBOP, based on a generalization of the Abell--Tersoff bond-order potential. By imposing physically motivated limiting constraints that enforce bond saturation, BOBOP exhibits robust extrapolation of potential energy surfaces to high-pressure and high-temperature conditions. Furthermore, the explicit high-order correlations embedded in the bonding-environment descriptor resolve bond-energy degeneracies, thereby improving the accuracy and transferability. Overall, comprehensive tests on molecular and condensed-phase systems demonstrate that BOBOP can serve as an efficient tool for reliable MD simulations and structure searching, offering insights into the design of inductive biases for controlling the extrapolation behavior of interatomic potentials.

\subsection{Limitations}
The current formulation of BOBOP also has several limitations. The symmetrization of edge features required to generate higher-order correlations becomes computationally demanding as the body order increases (see Appendix \ref{sec:efficiency}). In addition, the present framework does not account for long-range interactions and does not explicitly incorporate dihedral-angle dependence, which should modulate the bond order. Recursive evaluation of edge features, for example through message-passing architectures, may provide a route toward addressing these limitations. However, identifying a specific end-to-end functional form within such architectures that preserves both physical interpretability and robust extrapolation performance remains an open question for future work.

\section*{Data availability}
The implementation, training scripts, and optimized model parameters of BOBOP used to reproduce the results are provided at the GitHub repository \cite{bobop_github}. 
\section*{Acknowledgements}
I acknowledge the Center for Computational Materials Science, Institute for Materials Research, Tohoku University, for the use of the MASAMUNE-\RomanNumeralCaps{2} supercomputing facilities. This work was supported by the Japan Society for the Promotion of Science (KAKENHI Grant No. JP25KJ1170). I thank Prof. Albert P.~Bart\'ok for his helpful explanation of the GAP silicon dataset.

\appendix

\section{LOW-RANK APPROXIMATION OF WEIGHT TENSORS}
\label{sec:lowrank}
In this work, to reduce the number of independent trainable parameters, the weight tensors $W^{(3)}$, $W^{(4)}$, and $W^{(5)}$ in Eq.~\eqref{eq:zeta} are represented by low-rank approximations as 
\begin{align}
W_{cnlp}^{(3)} &= \sum_{q_{3}}w^{(3)}_{cnq_{3}}v^{(3)}_{q_{3}lp}, \\ 
W_{\substack{cn_{1}n_{2}\\l_{1}l_{2}l_{3}p}}^{(4)} &= \sum_{q_{4}}w^{(4)}_{cn_{1}n_{2}q_{4}}v^{(4)}_{q_{4}l_{1}l_{2}l_{3}p}, \\
W_{\substack{cn_{1}n_{2}n_{3}\\l_{1}l_{2}l_{3}l_{4}l_{5}p}}^{(5)} &= \sum_{q_{5}}w^{(5)}_{cn_{1}n_{2}n_{3}q_{5}}v^{(5)}_{q_{5}l_{1}l_{2}l_{3}l_{4}l_{5}p},
\end{align}
where $q_{3}$, $q_{4}$, and $q_{5}$ index the latent dimensions of the low-rank decompositions of $W^{(3)}$, $W^{(4)}$, and $W^{(5)}$, respectively. The indices $q_{3}$, $q_{4}$, and $q_{5}$ take values $1,\ldots,q_{3,\mathrm{max}}$, $1,\ldots,q_{4,\mathrm{max}}$, and $1,\ldots,q_{5,\mathrm{max}}$, respectively.

\section{HYPERPARAMETER SETTINGS AND TRAINING DETAILS}
\label{sec:hypara}

Hyperparameter settings of BOBOP are summarized in Table \ref{table:hyperparameter}. 
BOBOP was implemented in PyTorch \cite{3454287.3455008}. All trainable parameters were initialized to random small values from a uniform distribution and optimized with the mini-batch gradient descent algorithm. For parameter optimization, the AdamW \cite{loshchilov2019decoupledweightdecayregularization} optimizer was used with the ReduceLROnPlateau learning rate scheduler from PyTorch. The relative energy-loss weight $\rho$ was initially set to a small value of 0.1--10 to stabilize the training, and was subsequently increased to 10--1000 depending on the dataset.

\section{COMPUTATIONAL EFFICIENCY}
\label{sec:efficiency}

The computational efficiency of BOBOP was evaluated by timing MD simulations and compared with that of other representative local MLIP: GAP, DeePMD, and ACE. Timing measurements were performed using a single CPU core of an Intel Xeon Platinum 8480+ at 2.0 GHz. For BOBOP, the simulations were performed using ASE in combination with PyTorch's automatic differentiation functionality for force evaluation. For GAP, DeePMD, and ACE, LAMMPS \cite{LAMMPS} was used in combination with the \textsc{QUIP} \cite{Csanyi2007-py}, \textsc{DeePMD-kit} \cite{WANG2018178}, and \textsc{PACEMAKER} \cite{bochkarev2022efficient} packages, respectively. MD simulations were performed for 1000 silicon atoms at a density of 2.0 g/cm$^3$ and 512 water molecules at a density of 1.0 g/cm$^3$.

Table~\ref{table:timings} summarizes the inference times of the models. The three-body BOBOP achieves computational efficiency comparable to other models, whereas the computational cost increases substantially when four-body correlations are included. Thus, the current implementation and parameter settings of BOBOP are not necessarily computationally inexpensive, particularly when higher-order correlations are included. In this work, the same $l_{\max}$ was used for all correlation orders. However, using a smaller $l_{\max}$ for four-body and higher-order correlations may be a practical strategy for reducing the computational cost. Furthermore, in the present implementation, forces were evaluated through ASE using PyTorch's automatic differentiation functionality, which introduces some computational overhead. Explicitly implementing the analytical force expressions in LAMMPS or other molecular dynamics packages may improve computational efficiency. The comparison reflects the current implementations and should not be interpreted as an algorithmic comparison of optimal implementations.
\begin{widetext}
\section{EFFECTIVE BODY ORDER OF BOND-ORDER POTENTIALS}

\label{sec:relationship}

In this Appendix, we discuss the effective body order of bond-order potentials, Abell-Tersoff bond-order potential and BOBOP, by expanding them in ACE. 

\subsection{Effective body order of Abell-Tersoff bond-order potential}

We first discuss the effective body order of the Abell--Tersoff
bond-order potential. The many-body contribution of the Abell--Tersoff potential can be written as
\begin{align}
\label{eq:bondab}
E_{\mathrm{AB}}
&=\sum_{j\neq i} V_A(r_{ij})b_{ij}\notag\\
&=\sum_{j\neq i}\exp(-a r_{ij})\Phi(\zeta_{ij}),
\end{align}
where we assume an exponential attractive interaction
$V_A(r_{ij})=\exp(-a r_{ij})$ and use
$b_{ij}=\Phi(\zeta_{ij})$. Expanding the nonlinear function
$\Phi$ as a power series gives
\begin{align}
\label{eq:ps}
\Phi(\zeta_{ij})=\sum_{\nu=0}^{\nu_\mathrm{max}} C_{\nu} \zeta_{ij}^{\nu},
\end{align}
where $C_{\nu}$ is the expansion coefficient. The bond-environment descriptor
$\zeta_{ij}$ in Eq.~\eqref{eq:zeta_bo} can be expressed as
\begin{align}
\label{eq:zetaat}
\zeta_{ij}=\sum_{k\neq i,j}f(r_{ik})R(r_{ij})R'(r_{ik})\sum_l d_l \cos^l\theta_{ijk},
\end{align}
where we set $m=1$ and introduce $R(r_{ij})=\exp(\lambda r_{ij})$, $R'(r_{ik})=\exp(-\lambda r_{ik})$ and expand the angular function as $g(\theta_{ijk})=\sum_l d_l \cos^l\theta_{ijk}$.
Using Eqs.~\eqref{eq:ps}~and~\eqref{eq:zetaat}, Eq.~\eqref{eq:bondab} can be rewritten as
\begin{align}
E_{\mathrm{AB}}=\sum_{\nu=0}^{\nu_\mathrm{max}} C_{\nu} E_{\nu} ,
\end{align}
where
\begin{align}
E_{\nu}=&\sum_{j\neq i}\exp(-a r_{ij})\left[\sum_{k\neq i,j}f(r_{ik})R(r_{ij})R'(r_{ik})
\sum_{l} d_{l}\cos^l\theta_{ijk}\right]^{\nu} \notag\\
=&\sum_{j\neq i}\exp(-a r_{ij})\left[\sum_{k\neq i}f(r_{ik})R(r_{ij})R'(r_{ik})
\sum_{l} d_{l}\cos^l\theta_{ijk}-\sum_{l}d_{l}f(r_{ij})\right]^{\nu}.
\end{align}
The second term in the bracket, $\sum_{l}d_{l}f(r_{ij})$, originates from excluding the atom $j$ from the summation over neighboring atoms. Since this term only generates
lower-order contributions, we neglect it for simplicity and consider
\begin{align}
\label{eq:enu2}
\hat{E}_{\nu}=\sum_{j\neq i}\exp(-a r_{ij})\left[\sum_{k\neq
i}f(r_{ik})R(r_{ij})R'(r_{ik})\sum_{l}d_{l}\cos^l\theta_{ijk}\right]^{\nu} .
\end{align}

For $\nu=0$, this reduces to
\begin{align}
\hat{E}_{0} = \sum_{j \neq i} \exp(-a r_{ij}).
\end{align}

For $\nu=1$, 
\begin{align}
\label{eq:atnu1}
\hat{E}_{1} &= \sum_{j\neq i}\exp(-a r_{ij})\sum_{k \neq i} f(r_{ik})R(r_{ij})R^{\prime}(r_{ik})\sum_{l}d_{l}\cos^{l}\theta_{ijk} \notag \\
&= \sum_{j\neq i}\exp(-a r_{ij})\sum_{k \neq i} f(r_{ik})R(r_{ij})R^{\prime}(r_{ik})\sum_{l}d_{l}\sum_{\bm{l}}C(\bm{l})L_{\bm{l}}(\hat{\bm{r}}_{ij})L_{\bm{l}}(\hat{\bm{r}}_{ik}) \notag \\
&= \sum_{l}d_{l}\sum_{\bm{l}}C(\bm{l})\sum_{j\neq i}\exp(-a r_{ij})R(r_{ij})L_{\bm{l}}(\hat{\bm{r}}_{ij})\sum_{k \neq i}f(r_{ik})R^{\prime}(r_{ik})L_{\bm{l}}(\hat{\bm{r}}_{ik})  \notag \\
&= \sum_{l}d_{l}\sum_{\bm{l}}C(\bm{l})A_{i,\bm{l}}^{(1)}A^{\prime}_{i,\bm{l}},
\end{align}
where we have defined 
\begin{align}
A_{i,\bm{l}}^{(\nu)}=\sum_{j\neq i}\exp(-a r_{ij})\{R(r_{ij})\}^{\nu}L_{\bm{l}}(\hat{\bm{r}}_{ij}),
\end{align}
and 
\begin{align}
A^{\prime}_{i,\bm{l}}=\sum_{k \neq i}f(r_{ik})R^{\prime}(r_{ik})L_{\bm{l}}(\hat{\bm{r}}_{ik}). 
\end{align}
If we properly replace the radial-dependent factors by using appropriate basis functions, Eq.~\eqref{eq:atnu1} recovers the three-body contribution of ACE (in Cartesian form \cite{Cheng2024}).

For $\nu=2$, 
\begin{align}
\hat{E}_{2} =& \sum_{j\neq i}\exp(-a r_{ij})\left[\sum_{k\neq i}f(r_{ik})R(r_{ij})R'(r_{ik})\sum_{l}d_{l}\cos^l\theta_{ijk}\right]^2 \notag \\
=& \sum_{j\neq i}\exp(-a r_{ij})\sum_{k\neq i}f(r_{ik})R(r_{ij})R'(r_{ik})\sum_{l_{1}}d_{l_{1}}\cos^{l_{1}}\theta_{ijk}\sum_{h\neq i}f(r_{ih})R(r_{ij})R'(r_{ih})\sum_{l_{2}}d_{l_{2}}\cos^{l_{2}}\theta_{ijh} \notag \\
=& \sum_{j\neq i}\exp(-a r_{ij})\sum_{k\neq i}f(r_{ik})R(r_{ij})R'(r_{ik})\sum_{l_{1}}d_{l_{1}}\sum_{\bm{l}_{1}}C(\bm{l}_{1})L_{\bm{l}_{1}}(\hat{\bm{r}}_{ij})L_{\bm{l}_{1}}(\hat{\bm{r}}_{ik}) \notag \\ & \times \sum_{h\neq i}f(r_{ih})R(r_{ij})R'(r_{ih})\sum_{l_{2}}d_{l_{2}}\sum_{\bm{l}_{2}}C(\bm{l}_{2})L_{\bm{l}_{2}}(\hat{\bm{r}}_{ij})L_{\bm{l}_{2}}(\hat{\bm{r}}_{ih}) \notag \\
=& \sum_{l_{1}l_{2}}d_{l_{1}}d_{l_{2}}\sum_{\bm{l}_{1}\bm{l}_{2}}C(\bm{l}_{1})C(\bm{l}_{2})\sum_{j\neq i}\exp(-a r_{ij})R(r_{ij})^{2}L_{(\bm{l}_{1}+\bm{l}_{2})}(\hat{\bm{r}}_{ij})\sum_{k\neq i}f(r_{ik})R'(r_{ik})L_{\bm{l}_{1}}(\hat{\bm{r}}_{ik}) \sum_{h\neq i}f(r_{ih})R'(r_{ih})L_{\bm{l}_{2}}(\hat{\bm{r}}_{ih}) \notag \\
=& \sum_{l_{1}l_{2}}d_{l_{1}}d_{l_{2}}\sum_{\bm{l}_{1}\bm{l}_{2}}C(\bm{l}_{1})C(\bm{l}_{2})A^{(2)}_{i,(\bm{l}_{1}+\bm{l}_{2})}A^{\prime}_{i,\bm{l}_{1}}A^{\prime}_{i,\bm{l}_{2}}.
\end{align}
This expression corresponds to the four-body contribution, but the trainable coefficient tensor is restricted to a rank-one decomposed form $d_{l_{1}}d_{l_{2}}$. 

For $\nu \geq 1$, $\hat{E}_{\nu}$ can generally be expressed as
\begin{align}
\label{eq:gennu}
\hat{E}_{\nu}
&=
\sum_{l_1\cdots l_{\nu}}d_{l_1}\cdots d_{l_{\nu}}
\sum_{\bm l_1\cdots\bm l_{\nu}}C(\bm l_1)\cdots C(\bm l_{\nu})
A^{(\nu)}_{i,(\bm l_1+\cdots+\bm l_{\nu})}\prod_{s=1}^{\nu}A'_{i,\bm l_s},
\end{align}
which indicates that $E_{\nu}$ corresponds to a $(\nu+2)$-body correlation. Thus, the effective body order of the Abell--Tersoff bond-order potential is determined by the maximum polynomial degree $\nu_{\mathrm{max}}$, with the highest effective body order given by $\nu_{\mathrm{max}}+2$.

\subsection{Effective body order of BOBOP}
As for the Abell--Tersoff bond-order potential, we can similarly discuss the effective body order of BOBOP. To simplify the discussion, we consider the three-body bonding environment descriptor and restrict our discussion to $p_{\max}=1$. In this case, the bonding environment descriptor $\bm{\zeta}_{ij}$ has a single scalar component
\begin{align}
\label{eq:bobopzetaoned}
\zeta_{ij}
&=\sum_{cnl}
T_{c\mu_i\mu_j}W_{cnl}^{(3)}
\sum_{k\neq i,j}
\chi_{n\mu_{i}\mu_{k}}f(r_{ik})
R_{n}(r_{ij})
R'_{n}(r_{ik})
\cos^l\theta_{ijk},
\end{align}
where we have suppressed the index $p$ in Eq.~\eqref{eq:zeta} and defined $R_{n}(r_{ij}) = \exp(\lambda_{n}r_{ij})$ and $R'_{n}(r_{ik}) = \exp(-\lambda_{n}r_{ik})$. 
The many-body energy contribution of BOBOP can be written as
\begin{align}
\label{eq:bobop}
E_{\mathrm{BO}}
&=\sum_{j\neq i}f(r_{ij})D_{\mu_{i}\mu_{j}}^{(A)}\exp(-\beta_{\mu_i\mu_j}r_{ij})\Phi(\zeta_{ij}).
\end{align}
Expanding the nonlinear function $\Phi$ as a power series gives
\begin{align}
\label{eq:bobopthreepow}
\Phi(\zeta_{ij})&=\sum_{\nu=0}^{\nu_\mathrm{max}} C_{\nu} \zeta_{ij}^{\nu},
\end{align}
where $C_{\nu}$ is the expansion coefficient.
Using Eqs.~\eqref{eq:bobopzetaoned}~and~\eqref{eq:bobopthreepow}, Eq.~\eqref{eq:bobop} can be rewritten as
\begin{align}
E_{\mathrm{BO}}
=
\sum_{\nu=0}^{\nu_\mathrm{max}} C_{\nu} E_{\nu},
\end{align}
where
\begin{align}
E_{\nu}=&\sum_{j\neq i}f(r_{ij})D_{\mu_{i}\mu_{j}}^{(A)}\exp(-\beta_{\mu_i\mu_j}r_{ij})\left[\sum_{cnl}T_{c\mu_{i}\mu_{j}}W_{cnl}^{(3)}\sum_{k\neq i,j}
\chi_{n\mu_{i}\mu_{k}}f(r_{ik})
R_{n}(r_{ij})R'_{n}(r_{ik}) \cos^l\theta_{ijk}\right]^{\nu} \notag\\
=&\sum_{j\neq i}f(r_{ij})D_{\mu_{i}\mu_{j}}^{(A)}\exp(-\beta_{\mu_i\mu_j}r_{ij}) \notag \\
& \qquad \times\left[\sum_{cnl}T_{c\mu_{i}\mu_{j}}W_{cnl}^{(3)}\sum_{k\neq i}
\chi_{n\mu_{i}\mu_{k}}f(r_{ik})
R_{n}(r_{ij})R'_{n}(r_{ik}) \cos^l\theta_{ijk}-f(r_{ij})\sum_{cnl}T_{c\mu_{i}\mu_{j}}\chi_{n\mu_{i}\mu_{j}}W_{cnl}^{(3)}\right]^{\nu}.
\end{align}
The second term in the bracket, $f_c(r_{ij})T_{c\mu_{i}\mu_{j}}\sum_{cnl}W_{cnl}^{(3)}$, originates from excluding the atom $j$ from the summation over neighboring atoms. Since this term only generates lower-order contributions, we neglect it for simplicity and consider
\begin{align}
\label{eq:neu}
\hat{E}_{\nu}
=&\sum_{j\neq i}f(r_{ij})D_{\mu_{i}\mu_{j}}^{(A)}\exp(-\beta_{\mu_i\mu_j}r_{ij})\left[\sum_{cnl}T_{c\mu_{i}\mu_{j}}W_{cnl}^{(3)}\sum_{k\neq i}
\chi_{n\mu_{i}\mu_{k}}f(r_{ik})
R_{n}(r_{ij})R'_{n}(r_{ik}) \cos^l\theta_{ijk}\right]^{\nu}.
\end{align}
For later convenience, we define
\begin{align}
A_{i,c_{1} \cdots c_{\nu} n_{1} \cdots n_{\nu} \bm{l}_{1}\cdots\bm{l}_{\nu}}^{(\nu)} &= \sum_{j\neq i}f(r_{ij})D_{\mu_{i}\mu_{j}}^{(A)}\exp(-\beta_{\mu_i\mu_j}r_{ij})\left(\prod_{s=1}^{\nu}T_{c_{s}\mu_{i}\mu_{j}}R_{n_{s}}(r_{ij})L_{\bm{l}_{s}}(\hat{\bm{r}}_{ij})\right), \notag \\
A'_{i,n\bm{l}} &= \sum_{k\neq i}\chi_{n\mu_{i}\mu_{k}}f(r_{ik})R'_{n}(r_{ik})L_{\bm{l}}(\hat{\bm{r}}_{ik}).
\end{align}

For $\nu=0$, $\hat{E}_{\nu}$ reduces to
\begin{align}
\hat{E}_{0} &= \sum_{j\neq i}f(r_{ij})D_{\mu_{i}\mu_{j}}^{(A)}\exp(-\beta_{\mu_i\mu_j}r_{ij}).
\end{align}

For $\nu \geq 1$, $\hat{E}_{\nu}$ can generally be expressed as 
\begin{align}
\hat{E}_{\nu} =& \sum_{\substack{c_{1} \cdots c_{\nu}\\n_{1} \cdots n_{\nu}\\l_{1} \cdots l_{\nu}}}W_{c_{1}n_{1}l_{1}}^{(3)}\cdots W_{c_{\nu}n_{\nu}l_{\nu}}^{(3)}
\sum_{\bm l_1\cdots\bm l_{\nu}}C(\bm l_1)\cdots C(\bm l_{\nu})A_{i,c_{1} \cdots c_{\nu} n_{1} \cdots n_{\nu} \bm{l}_{1}\cdots\bm{l}_{\nu}}^{(\nu)}\left(\prod_{s=1}^{\nu}A'_{i,n_{s}\bm{l}_{s}}\right).
\end{align}
This expression is an extended form of Eq.~\eqref{eq:gennu} that incorporates multiple radial bases and an additional dimension for chemical encoding, and $E_{\nu}$ corresponds to a $(\nu+2)$-body correlation. As in the Abell--Tersoff bond-order potential, the effective body order of BOBOP depends on the nonlinearity of the function $\Phi$. 

In this discussion, we set $p_{\max}=1$, such that the weight tensors corresponding to the implicit high-order contributions are represented by tensor products of the weight tensors for the three-body contributions $W^{(3)}$ in a rank-one decomposed form. Increasing $p_{\max}$ to two or more relaxes the rank-one constraint, allowing the weight tensors to be represented by a sum of multiple tensor products and thereby providing a more flexible approximation of the high-rank weight tensors for the implicit high-order contributions.
\end{widetext}

\bibliography{apssamp}

\end{document}